\documentclass[preprint,twocolumn]{aastex6}
\usepackage{graphicx}
\usepackage{amsmath}	
\usepackage{amssymb}	
\usepackage{multirow}
\usepackage{xfrac}
\usepackage{subfigure}
\usepackage{subfloat}

\graphicspath{{./} }

\begin{document}

\title{IC 3639 -- A new bona fide Compton thick AGN unveiled by \textit{NuSTAR}}

\author{Peter G. Boorman,\altaffilmark{1} P. Gandhi,\altaffilmark{1} D. Alexander,\altaffilmark{2} A. Annuar,\altaffilmark{2}  D. R. Ballantyne,\altaffilmark{3} F. Bauer,\altaffilmark{4,5,6} S. E. Boggs,\altaffilmark{7} W. N. Brandt,\altaffilmark{8,9,10} M. Brightman,\altaffilmark{11} F. E. Christensen,\altaffilmark{12} W. W. Craig,\altaffilmark{12,13} D. Farrah,\altaffilmark{14} C. J. Hailey,\altaffilmark{15} F. A. Harrison,\altaffilmark{13} S. F. H{\"o}nig,\altaffilmark{1} M. Koss,\altaffilmark{16} S. M. LaMassa,\altaffilmark{17} A. Masini,\altaffilmark{18,19} C. Ricci,\altaffilmark{4} G. Risaliti,\altaffilmark{20} D. Stern,\altaffilmark{21} W. W. Zhang\altaffilmark{22}}

\email{p.g.boorman@soton.ac.uk}
\altaffiltext{1}{Department of Physics \& Astronomy, Faculty of Physical Sciences and Engineering, University of Southampton, Southampton, SO17 1BJ, UK; p.g.boorman@soton.ac.uk}
\altaffiltext{2}{Centre for Extragalactic Astronomy, Department of Physics, Durham University, South Road, Durham, DH1 3LE, UK}
\altaffiltext{3}{Center for Relativistic Astrophysics, School of Physics, Georgia Institute of Technology, Atlanta, GA 30332, USA}
\altaffiltext{4}{Instituto de Astrof{\'{\i}}sica and Centro de Astroingenier{\'{\i}}a, Facultad de F{\'{i}}sica, Pontificia Universidad Cat{\'{o}}lica de Chile, Casilla 306, Santiago 22, Chile}
\altaffiltext{5}{Millennium Institute of Astrophysics (MAS), Nuncio Monse{\~{n}}or S{\'{o}}tero Sanz 100, Providencia, Santiago, Chile}
\altaffiltext{6}{Space Science Institute, 4750 Walnut Street, Suite 205, Boulder, Colorado 80301}
\altaffiltext{7}{Space Sciences Laboratory,University of California, Berkeley, CA 94720, USA}
\altaffiltext{8}{Department of Astronomy and Astrophysics, The Pennsylvania State University, 525 Davey Lab, University Park, PA 16802, USA}
\altaffiltext{9}{Institute for Gravitation and the Cosmos, The Pennsylvania State University, University Park, PA 16802, USA}
\altaffiltext{10}{Department of Physics, 104 Davey Lab, The Pennsylvania State University, University Park, PA 16802, USA}
\altaffiltext{11}{Cahill Center for Astronomy and Astrophysics, California Institute of Technology, Pasadena, CA 91125, USA}
\altaffiltext{12}{DTU Space-National Space Institute, Technical University of Denmark, Elektrovej 327, DK-2800 Lyngby, Denmark}
\altaffiltext{13}{Lawrence Livermore National Laboratory, Livermore, CA 94550, USA}
\altaffiltext{14}{Department of Physics, Virginia Tech, Blacksburg, VA 24061, USA}
\altaffiltext{15}{Columbia Astrophysics Laboratory, Columbia University, New York, NY 10027, USA}
\altaffiltext{16}{Institute for Astronomy, Department of Physics, ETH Zurich, Wolfgang-Pauli-Strasse 27, CH-8093 Zurich, Switzerland}
\altaffiltext{17}{NASA Postdoctoral Program Fellow, NASA Goddard Space Flight Center, Code 665, Greenbelt, MD 20771, USA}
\altaffiltext{18}{INAF-Osservatorio Astronomico di Bologna, via Ranzani 1, 40127 Bologna, Italy}
\altaffiltext{19}{Dipartimento di Fisica e Astronomia (DIFA), Universit\`a di Bologna, viale Berti Pichat 6/2, 40127 Bologna, Italy}
\altaffiltext{20}{INAF–Arcetri Observatory, Largo E. Fermi 5, 50126 Firenze, Italy}
\altaffiltext{21}{Jet Propulsion Laboratory, California Institute of Technology, Pasadena, CA 91109, USA}
\altaffiltext{22}{X-ray Astrophysics Laboratory, NASA Goddard Space Flight Center, Greenbelt, MD 20771, USA}

\begin{abstract}
We analyse high-quality \textit{NuSTAR} observations of the local (\textit{z}\,=\,0.011) Seyfert 2 active galactic nucleus (AGN) IC 3639, in conjunction with archival \textit{Suzaku} and \textit{Chandra} data.  This provides the first broadband X-ray spectral analysis of the source, spanning nearly two decades in energy (0.5\,--\,30\,keV).  Previous X-ray observations of the source below 10\,keV indicated strong reflection/obscuration on the basis of a pronounced iron fluorescence line at 6.4\,keV.  The hard X-ray energy coverage of \textit{NuSTAR}, together with self-consistent toroidal reprocessing models, enables direct broadband constraints on the obscuring column density of the source.  We find the source to be heavily Compton-thick (CTK) with an obscuring column in excess of $3.6\times10^{24}$\,cm$^{-2}$, unconstrained at the upper end. We further find an intrinsic 2\,--\,10\,keV luminosity of $\textrm{log}_{10}(L_{\textrm{2\,--\,10\,keV}}\,\textrm{[erg\,s}^{-1}])\,=\,43.4^{+0.6}_{-1.1}$ to 90\% confidence, almost 400 times the observed flux, and consistent with various multi-wavelength diagnostics.  Such a high intrinsic to observed flux ratio in addition to an Fe-K$\alpha$ fluorescence line equivalent width exceeding 2\,keV is extreme amongst known \textit{bona fide} CTK AGN, which we suggest are both due to the high level of obscuration present around IC 3639.  Our study demonstrates that broadband spectroscopic modelling with \textit{NuSTAR} enables large corrections for obscuration to be carried out robustly, and emphasises the need for improved modelling of AGN tori showing intense iron fluorescence.
\end{abstract}


\section{INTRODUCTION}
\label{sec:introduction}

The origin of the cosmic X-ray background (CXB) has been under study ever since its discovery more than 60 years ago \citep{Giacconi1962}.  Spanning from fractions of a keV (soft X-rays) up to several hundreds of keV (hard X-rays), the general consensus today is that the majority of the CXB arises from the integrated emission of discrete sources of radiation, with the most prominent contribution arising from active galactic nuclei (AGN) (e.g. \citealt{Mushotzky2000}).  The unified model of AGN \citep{Antonucci1993,Netzer2015} predicts that the major differences seen between different classes of AGN can be attributed to an orientation effect, with the primary radiation source being surrounded by an obscuring torus inclined relative to our line-of-sight (LOS).  This leads to effectively two types of AGN - those with a direct view to the nucleus (largely unobscured) and those with an obscured view to the nucleus from behind a putative torus (see \citealt{Marin2016} for a recent review on the orientation of AGN).  In addition, obscured AGN have been required to fit the CXB, with \citet{Setti1989} requiring a considerable contribution from this AGN population.  Multiple studies have revealed this to be the case, although with a dependence on X-ray luminosity (e.g. \citealt{Lawrence2010}).  This suggests heavily obscured AGN may be a major contributor to the CXB, and there are many ongoing efforts to study this population (e.g. \citealt{Brandt2015}, and references therein).

In the X-ray band, the two important interaction processes between photons and matter surrounding an AGN are photoelectric absorption and Compton scattering.  Photoelectric absorption is dominant at lower energies, whereas Compton scattering dominates in hard X-rays above $\sim$10\,keV up to the Klein-Nishina decline.  X-ray photons with energy greater than a few keV are visible if the LOS obscuring column density ($N_{\textrm H}$) is $\lesssim$\,1.5\,$\times\,10^{24}\,\textnormal{cm}^{-2}$, and such AGN are named \textit{Compton-thin} (CTN) since the matter is optically thin to Compton scattering and a significant fraction of the photons with \textit{E}\,\textgreater\,10\,keV escape after one or more scatterings.  This leads to only slight depletion of hard X-rays for CTN sources.  Sources with column densities greater than this value are classified as \textit{Compton-thick} (CTK) since even high-energy X-rays \textit{can} be diminished via Compton scattering, leading to the X-ray spectrum being depressed over the entire energy range.  The hard X-ray spectrum of \textit{typical} CTK AGN is characterised by three main components: a Compton reflection hump, peaking at $\sim$30\,keV; a strong neutral Fe-K$\alpha$ fluorescence line at $\sim$6.4\,keV \citep{Matt2000} (\textit{strong} generally refers to an equivalent width EW\,$\gtrsim$\,1\,keV); and an underlying absorbed power law with an upper cut-off of several hundred keV (intrinsic to the AGN, arising from the Comptonisation of accretion disc photons in the corona).  The ability to detect the absorbed power law in the spectrum of a source depends on the level of obscuration - in heavily CTK sources, this component is severely weakened and can be entirely undetectable.  The Compton hump and Fe-K$\alpha$ line are both reflection features from the putative torus.

X-ray selection is one of the most effective strategies available for detecting CTK sources because hard X-ray photons stand a greater chance of escaping the enshrouding obscuring media due to their increased penetrating power.  In addition, photons with initial propagation directions out of the LOS can also be detected through Compton scattering into our LOS.  For this reason, the best energy range to observe CTK AGN is \textit{E}\,\textgreater\,10\,keV.  In general, many synthesis models formulated to date seem to agree that fitting to the peak flux of the CXB at $\sim$30\,keV requires a CTK AGN contribution in the range 10\,--\,25\% \citep{Comastri1995,Gandhi2003,Gilli2007,Treister2009,Draper2010,Akylas2012,Ueda2014}.  The actual number density of CTK AGN remains unclear, with various recent sample observations suggesting a fraction exceeding 20\% \citep{Goulding2011,Lansbury2015,Ricci2015,Koss2016b}.  \citet{Gandhi2007} and \citet{Treister2009} discuss degeneracy between the different component parameters (e.g. reflection and obscuration) used to fit the CXB.  This is why the shape of the CXB cannot be directly used to determine the number of CTK AGN, and further explains the large uncertainty associated with the CTK fraction.

Many X-ray missions to date have been capable of detecting photons above 10\,keV, such as \textit{BeppoSAX}, \textit{Swift}, \textit{Suzaku} and \textit{INTEGRAL}.  However, due to issues including high background levels, relatively small effective areas and low angular resolution, few CTK sources have been identified.  The \textit{Nuclear Spectroscopic Telescope Array} (\textit{NuSTAR}) \citep{Harrison2013} is the first mission in orbit capable of \textit{true} X-ray imaging in the energy range $\sim$3\,--\,79\,keV.
Since launch, \textit{NuSTAR} has not only studied well known CTK AGN in detail \citep{Arevalo2014,Bauer2015,Marinucci2016}, it has also helped to identify and confirm numerous CTK candidates in the local Universe \citep{Gandhi2014,Balokovic2014,Annuar2015,Koss15,Koss2016}, as well as carry out variability studies focusing on \textit{changing-look} AGN \citep{Risaliti2013,Walton2014,Rivers2015,Marinucci2016,Ricci2016,Masini2016b}.  Moreover, deep \textit{NuSTAR} surveys have resolved a fraction of 35$\pm$5\% of the total integrated flux of the CXB in the 8\,--\,24\,keV band \citep{Harrison2015}.

Detailed modelling of individual highly obscured sources is the most effective way to understand the spectral components contributing to the missing fraction of the peak CXB flux.
Here we carry out the first robust broad-band X-ray spectral analysis of the nearby Seyfert 2 and candidate CTK AGN IC 3639 (also called Tololo 1238-364). The source is hosted by a barred spiral galaxy (Hubble classification SBbc\footnote{http://leda.univ-lyon1.fr}) with redshift \textit{z}\,=\,0.011 and corresponding luminosity distance D\,=\,53.6\,Mpc.  This is calculated for a flat cosmology with $H_{0}$\,=\,67.3 \,km\,s$^{-1}$\,Mpc$^{-1}$, $\Omega_{\Lambda}$\,=\,0.685 and $\Omega_{M}$\,=\,0.315 \citep{Planck2014}.  All uncertainties are quoted at a 90\% confidence level for one interesting parameter, unless stated otherwise.
This paper uses \textit{NuSTAR} and archival X-ray data from the \textit{Suzaku} and \textit{Chandra} satellites.  The \textit{Suzaku} satellite operated in the energy range $\sim$0.1\,--\,600\,keV and is thus capable of detecting hard X-rays.  However, the hard X-ray energy range of this satellite is covered by a non-imaging detector, leading to potential complications for faint sources, as outlined in Section \ref{sec:obs_SU_HXD}.  \textit{Chandra} has a high energy limit of $\sim$8\,keV, and very high angular resolution with a lower energy limit $\gtrsim$0.1\,keV.  Consequently, the different capabilities of \textit{NuSTAR}, \textit{Suzaku} and \textit{Chandra} complement each other so that a multi-instrument study provides a \textit{broad-band} spectral energy range.

The paper is structured accordingly: Section \ref{sec:target_selection} explains the target selection, with Section \ref{sec:observations} describing the details behind each X-ray observation of the source used as well as the spectral extraction processes.  The corresponding X-ray spectral fitting and results are outlined in Sections \ref{sec:x-ray_fitting} and \ref{sec:results}, respectively.  Finally, Section \ref{sec:discussion} outlines broadband spectral components determined from the fits, the intrinsic luminosity of the source and a multi-wavelength comparison with other CTK sources.  We conclude with a summary of our findings in Section \ref{sec:conclusions}.

\section{THE TARGET}
\label{sec:target_selection}

The first published X-ray data of IC 3639 were reported by \citet{Risaliti1999}, where they suggest the source to be CTK with column density $N_\textrm{H}$\,\textgreater\,$10^{25}\,\textnormal{cm}^{-2}$.  This lower limit was determined from a soft X-ray spectrum provided by the \textit{BeppoSAX} satellite, together with multi-wavelength diagnostic information (see \citealt{Risaliti1999b} and references therein for further details on the modelling used).  Additionally, the EW of the Fe-K$\alpha$ emission-line was reported as $3.20^{\,+0.98}_{\,-1.74}$\,keV.  Such high EWs are extreme though not unheard of, as reported by \citet{Levenson2002}.  Optical images of the source, as well as surrounding source redshifts, infer IC 3639 to be part of a triple merger system (e.g. Figure \ref{fig:1a}, upper left panel - IC 3639 is $\sim$1\farcm5 away from its nearest galaxy neighbour to the North-East).  However, \citet{Barnes2001} use the HI detection in this interacting group to suggest that it is free of any significant galaxy interaction or merger.

\citet{Miyazawa2009} analysed the source as part of a sample of 36 AGN observed by \textit{Suzaku}, including the higher energy HXD PIN data.  The source was found to have an obscuring column density of $7.47^{+4.81}_{-3.14}\,\times10^{23}\,\textnormal{cm}^{-2}$ with photon index $1.76^{+0.52}_{-0.44}$, suggesting a CTN nature.  This \textit{could} indicate variability between the 2007 \textit{Suzaku} observation and the 1999 \textit{BeppoSAX} observation reported by \citet{Risaliti1999}.

As outlined in Section \ref{sec:introduction}, CTK sources are notoriously hard to detect due to their low count rate in the soft X-ray band.  For this reason, one must use particular spectral characteristics, indicative of CTK sources.  The first, most obvious indication is a prominent Fe-K$\alpha$ fluorescence line.  This can occur if the fluorescing material is exposed to a greater X-ray flux than is directly observed, so that the line appears strong relative to the continuum emission \citep{Krolik1987}.

Other CTK diagnostics are provided through multi-wavelength analysis.  For example, by comparing the MIR and X-ray luminosities of the source, which have been shown to correlate for AGN \citep{Elvis1978,Krabbe2001,Horst2008,Gandhi2009,Levenson2009,Mateos2015,Stern2015,Asmus2015}.  X-ray obscuration is expected to significantly offset CTK sources from this relation.  Indeed, IC 3639 shows an \textit{observed} weak X-ray (2\,--\,10\,keV) luminosity compared to the \textit{predicted} value from this correlation.
Another multi-wavelength technique compares emission lines originating in the narrow line region (NLR) on larger scales than the X-ray emission, which arises close to the core of the AGN.  Of the multitude of emission lines available for such analysis, one well studied correlation uses the optical [OIII] emission-line flux at $\lambda$\,=\,5007\AA.  \citet{Panessa2006} and \citet{Berney2015}, among others, study a correlation between the observed [OIII] emission-line luminosity and X-ray (2\,--\,10\,keV) luminosity for a group of Seyfert galaxies, after correcting for obscuration.  IC 3639 again shows a weak X-ray flux compared to the observed [OIII] luminosity.  This indicates heavy obscuration depleting the X-ray luminosity.  For a comparison between the ratios of MIR and [OIII] emission flux to X-ray flux with the average value for (largely unobscured) Seyfert 1s, see \citet[Figure 2]{LaMassa2010}.

\citet{Dadin2007} reports that the \textit{BeppoSAX} observation of IC 3639 in both the 20\,--\,100\,keV and 20\,--\,50\,keV bands had negligible detection significance, and place an upper bound on the 20\,--\,100\,keV flux of $F_{\textnormal{20\,--\,100\,keV}}\leqslant$\,9.12\,$\times$\,10$^{-12}$\,erg\,s$^{-1}$\,cm$^{-2}$.  Unfortunately, IC 3639 lies below the \textit{Swift}/BAT all-sky survey limit of $\sim$1.3$\,\times\,10^{-11}$\,erg\,s$^{-1}$\,cm$^{-2}$ in the 14\,--\,195\,keV band \citep{Baumgartner2013}.

\begin{subfigures}
\begin{figure*}
  \includegraphics[width = 1\textwidth]{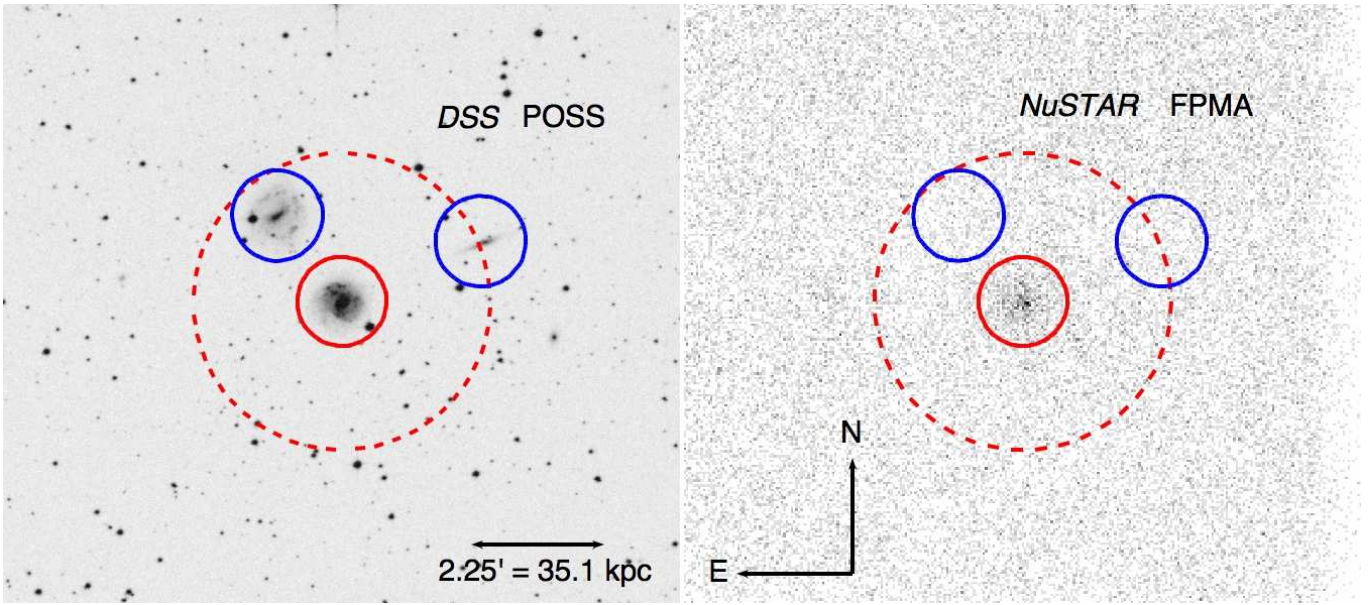}
  \caption{\label{fig:1a}(Left) \textit{Digitised Sky Survey} (\textit{DSS}) optical image of the interacting triple galaxy group system, with IC 3639 shown in the center.  The \textit{NuSTAR} extraction regions for source and background are shown in red with solid and dashed lines respectively.  The other two galaxies present in the triple system are highlighted with blue circles, each of radius 0\farcm76.  (Right) \textit{NuSTAR} FPMA event file with source and background regions again shown in red solid and dashed lines respectively.  The locations of the other two galaxies in the interacting triple system (visible in optical images) are shown by the blue circles.  Clearly, \textit{NuSTAR} does not significantly detect these other galaxies in the X-ray energy range.}
\end{figure*}

\begin{figure*}
  \includegraphics[width = 1\textwidth]{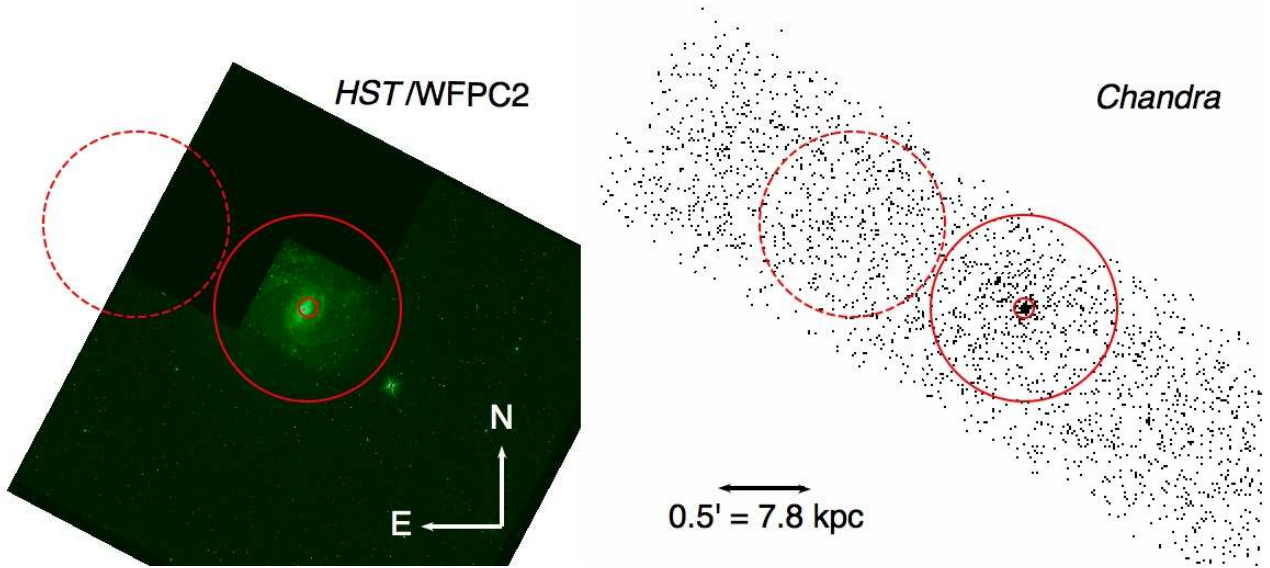}
  \caption{\label{fig:1b}(Left) \textit{Hubble Space Telescope} (\textit{HST}) WFPC2 image of IC 3639, with superimposed regions defined from the \textit{Chandra} image.  The image confirms the barred spiral classification described in Section \ref{sec:introduction}, orientated almost completely face-on to our LOS.  (Right)  Full band \textit{Chandra} image with 0\farcm5 radius extraction regions for background and annular count extraction region for off-nuclear emission superimposed.  The annular extraction region has an inner radius of 0\farcm05.  Scale was determined assuming a distance to the source of 53.6\,Mpc.}
\end{figure*}
\end{subfigures}

\section{OBSERVATIONS \& DATA REDUCTION}
\label{sec:observations}

Archival observations for IC 3639 used in this paper were all extracted from the HEASARC archive\footnote{https://heasarc.gsfc.nasa.gov/cgi-bin/W3Browse/w3browse.pl}.  Together, we use \textit{Suzaku} (XIS \& HXD), \textit{Chandra} and recent \textit{NuSTAR} data in this study.  Table \ref{tab:obs_info} shows the details of each of these observations.

 \begin{table}[h]
 	\begin{center}
 	\begin{tabular}{rllll}
  		\hline \hline
 		Satellite 					& Obs. ID 							& Date \,/\,\textsc{y-m-d} 	& Exp.\,/\,ks & PI\\
 		\hline \hline
 		\multicolumn{1}{r}{\textit{NuSTAR}} & \multicolumn{1}{l}{60001164002}  	& 2015-01-09 				& \multicolumn{1}{l}{58.7} & P. Gandhi\\
 		\hline
 		\multicolumn{1}{r}{\textit{Suzaku}} & \multicolumn{1}{l}{702011010} 	& 2007-07-12 				& \multicolumn{1}{l}{53.4} & H. Awaki\\
 		\hline
 		\multicolumn{1}{r}{\textit{Chandra}} & \multicolumn{1}{l}{4844}		 	& 2004-03-07 				& \multicolumn{1}{l}{9.6} & R. Pogge\\
 		\hline
 		\multicolumn{5}{p{.5\textwidth}}{Col. 1: Satellite name; Col. 2: Corresponding observation ID; Col. 3: Date of observation; Col. 4: Unfiltered exposure time for the observation; Col. 5: Principal Investigator (PI) of the observation.}
 	\end{tabular}
 	\end{center}
 	\caption{Details of observations used to analyse IC 3639.}
 	\label{tab:obs_info}
\end{table}

\subsection{\textit{NuSTAR}}
\label{sec:obs_NU}

Data from both focal plane modules (FPMA \& FPMB) onboard the \textit{NuSTAR} satellite were processed using the \textit{NuSTAR} Data Analysis Software (\textsc{NuSTARDAS}) within the \textsc{heasoft} package.  The corresponding \textsc{caldb} files were used with the \textsc{NuSTARDAS} task \textsc{nupipeline} to produce calibrated and cleaned event files.  The spectra and response files were produced using the \textsc{nuproducts} task, after standard data screening procedures.  The net count rate in the 3\,--\,79\,keV band for FPMA \& FPMB were $(9.413\,\pm\,0.549)\times\,10^{-3}$\,counts\,s$^{-1}$ and $(8.018\,\pm\,0.544)\times\,10^{-3}$\,counts\,s$^{-1}$, for net exposures of 58.7\,ks and 58.6\,ks, respectively (this corresponds to total count rates of $(1.686\,\pm\,0.054)\times\,10^{-2}$\,counts\,s$^{-1}$ and $(1.648\,\pm\,0.053)\times\,10^{-2}$\,counts\,s$^{-1}$ for FPMA \& FPMB, respectively).
Circular source regions of radius 0\farcm75 were used to extract source counts from the corresponding event files.  Background counts were extracted from annular regions of outer radius 2\farcm5 and inner radius 0\farcm75, centred on the source regions to avoid any cross-contamination between source and background counts.  The background region was chosen to be as large as possible in the same module as the source.

\begin{figure*}
\centering
\includegraphics[angle=-90,width=0.8\textwidth,trim={0cm 74cm 8cm 0cm},clip]{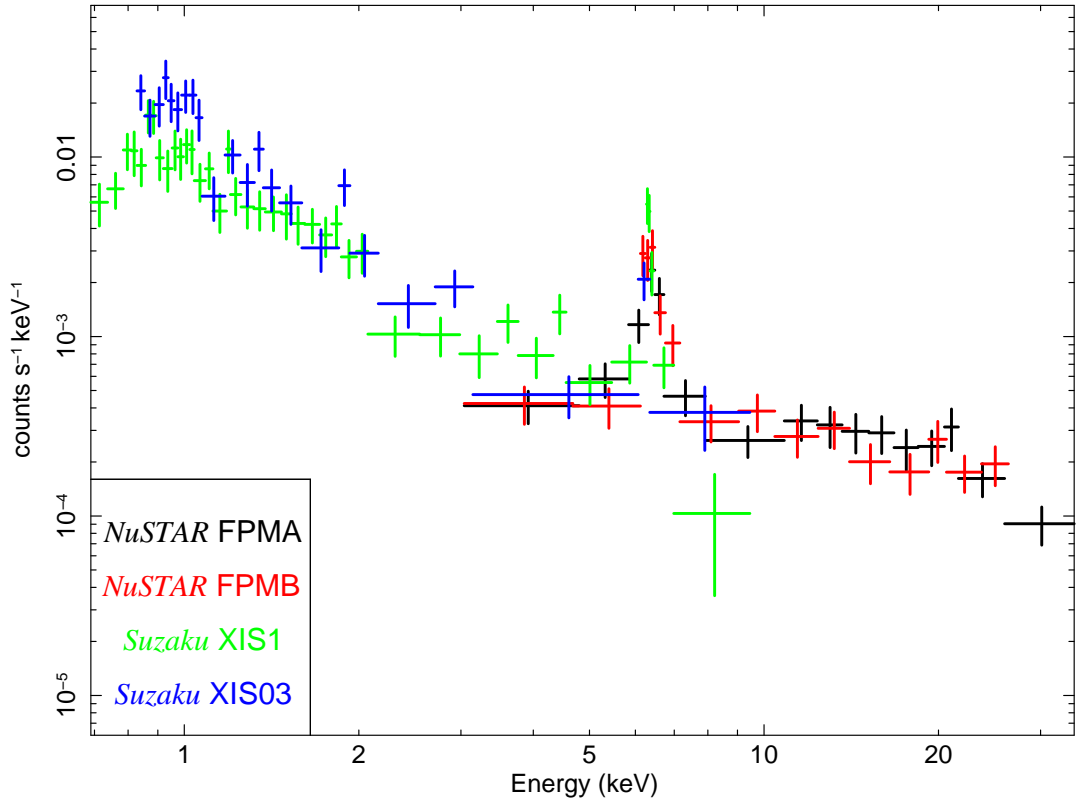}
\caption{The \textit{NuSTAR} FPMA and FPMB spectra plotted together with the \textit{Suzaku} XIS spectra in count rate units.  All are binned with SNR of 4. The data has been normalised by the area scaling factor present in each response file - this was unity for each observation.  XIS1 refers to the single BI detector on \textit{Suzaku} used to collect the spectral counts, whereas XIS03 refers to the combined spectra from the two FI detectors, XIS0 and XIS1 - see Section \ref{sec:obs_SU_XIS} for further details.}
\label{fig:f2}
\end{figure*}

The extracted spectra for FPMA and FPMB were then analysed using the  \textsc{xspec} version 12.9.0 software package\footnote{https://heasarc.gsfc.nasa.gov/xanadu/xspec/XspecManual.pdf}.  The energy range was constrained to the optimum energy range of \textit{NuSTAR} and grouped so that each bin contained a signal-to-noise ratio (SNR) of at least 4.  The resulting spectra are shown in count-rate units in Figure \ref{fig:f2}.

Figure \ref{fig:1a} shows the comparison of the \textit{NuSTAR} FPMA image with an optical \textit{Digitised Sky Survey} (\textit{DSS}) image.  The blue regions highlight the counterparts of the merging triple, clearly visible in the optical.  However, there is no detection of the separate galaxies in the \textit{NuSTAR} image, with the primary emission originating from IC 3639.

\subsection{\textit{Suzaku}}
\label{sec:obs_SU}

When fully operational, \textit{Suzaku} had four CCD X-ray imaging spectrometers (XIS) and a hard X-ray detector (HXD).  The XIS covered an energy range of 0.4\,--\,10\,keV with typical resolution 120\,eV.\footnote{isas.jaxa.jp/e/enterp/missions/suzaku/index.shtml}  During the lifetime of \textit{Suzaku}, one of the four XIS detectors became non-operational, leaving two front illuminated (FI) detectors (XIS0 and XIS3), and one back illuminated (BI) detector (XIS1).
HXD was a non-imaging instrument designed for observations in the energy range 10\,--\,700\,keV.

\subsubsection{XIS}
\label{sec:obs_SU_XIS}
First, the \textsc{ximage} software package\footnote{https://heasarc.gsfc.nasa.gov/xanadu/ximage/ximage.html} was used to create an image by summing over the three XIS cleaned event files.  Next source counts were extracted from a circular region of radius 2\farcm6, with background counts extracted from an annular region of inner radius 2\farcm6 and outer radius 5\farcm0.  The background annular region was again centred around the source region to avoid source and background count contamination.
\textsc{xselect} was then used to extract a spectrum for each XIS detector cleaned event file using the source and background regions defined above.  Lastly, we used the \textsc{addascaspec} command to combine the two FI XIS spectra.
The final result was two spectra: one for the FI cameras (XIS0 + XIS3, referred to as XIS03 herein) and one for the single BI camera (XIS1).  The net exposure times for XIS03 and XIS1 were 107.8\,ks and 53.4\,ks, respectively.  The data were again grouped with a minimum SNR of 4.  Additionally the XIS spectral data in the energy range 1.7\,--\,1.9\,keV and 2.1\,--\,2.3\,keV were ignored due to instrumental calibration uncertainties associated with the silicon and gold edges in the spectra\footnote{http://heasarc.gsfc.nasa.gov/docs/suzaku/analysis/abc/node8.html}.

\subsubsection{HXD}
\label{sec:obs_SU_HXD}
\begin{figure}
\center
\includegraphics[angle=-90,width=1\columnwidth]{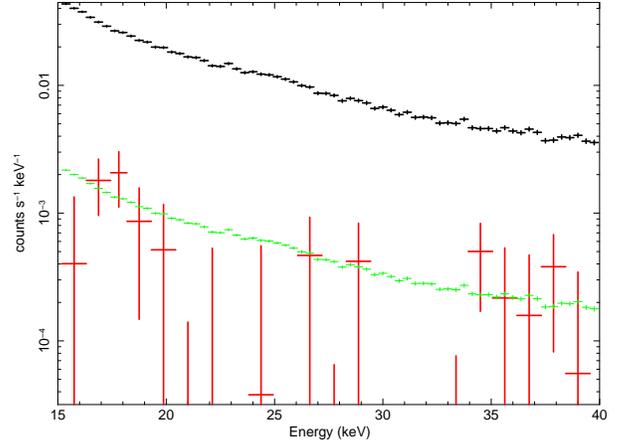}
\caption{The \textit{Suzaku} HXD spectra plotted in count rate units for gross counts (source\,+\,background), 5\% gross counts (to compare the approximate threshold we use to classify a weak detection - see text for details) and net source counts shown in black, green and red, respectively.  The HXD 15\,--\,40\,keV spectral counts are shown not to be significantly detected by \textit{Suzaku}, as discussed in Section \ref{sec:obs_SU_HXD}.}
\label{fig:f3}
\end{figure}

The corresponding spectrum for the HXD instrument was generated with the \textsc{ftools} command $\tt hxdpinxbpi$.
The data were then binned to allow a minimum of 500 counts per bin. The energy range 10\,--\,700\,keV of HXD is achieved with Gadolinium Silicate (GSO) counters for \textgreater\,50\,keV and PIN diodes for the range 15\,--\,50\,keV.  The GSO instrument is significantly less sensitive than \textit{NuSTAR} and thus not used here.  For the PIN instrument, a model has been designed to simulate the non-X-ray background (NXB).  In the 15\,--\,40\,keV range, current systematic uncertainties in the modelled NXB are estimated to be $\sim$3.2\%.  A \textit{tuned} NXB file for the particle background is provided by the \textit{Suzaku} team, whereas the CXB is evaluated separately and added to the tuned background resulting in a final \textit{total} background.  The modelled CXB is $\sim$5\% of the total background for PIN.  The \textsc{ftool} command \texttt{hxdpinxbpi} then uses the total background to produce a dead-time corrected PIN source and background (NXB\,+\,CXB) spectrum.  The net source counts for IC 3639 are shown in red in Figure \ref{fig:f3}.  The gross counts (source\,+\,background) are considerably higher than the net source counts and are shown in black on the same figure for comparison.  The source flux was calculated in the energy range 10\,--\,40\,keV for a simple power-law model.  The corresponding fluxes for source\,+\,background ($F_{\textrm{B,15-40\,keV}}$) and source alone ($F_{\textrm{S,15-40\,keV}}$) are:\\

\noindent $F_{\textrm{B,15-40\,keV}}=1.41\,\pm\,0.01\,\times\,10^{-10}$ erg\,s$^{-1}$\,cm$^{-2}$, and\\
$F_{\textrm{S,15-40\,keV}}=8.20\,^{+0.47}_{-8.20}\,\times\,10^{-13}$ erg\,s$^{-1}$\,cm$^{-2}$.\\

The CXB is known to vary between different instruments on the order of $\sim$\,10\% in the energy range considered here.  For this reason, as a consistency check, we compared the background uncertainty from \textit{Suzaku} ($3-5\%$)\footnote{http://heasarc.gsfc.nasa.gov/docs/suzaku/analysis/abc/node10.html, §7.5.1} to the error in the total background found when the CXB flux component carried a $10\%$ uncertainty.  This altered the tuned background error to 2.9\%\,--\,4.8\%.  Thus, within acceptable precision, the total background appears to be unaltered by potential CXB cross-instrument fluctuations.
If the source spectrum is less than $\sim$5\% of the tuned background, the detection is weak, and a source spectrum flux lower than $\sim$3\% of the background would require careful assessment.  The IC 3639 source counts are found to be $0.8\,^{+0.8}_{-0.9}\%$ of the tuned background counts in the 15\,--\,40\,keV range.  For this reason, we do not use the HXD data in our spectral analysis for IC 3639.  This value contradicts \citet{Miyazawa2009} who report a 15\,--\,50\,keV flux of $F_{15-50\,keV}\,=\,1.0\,\times\,10^{-11}$\,erg\,s$^{-1}$\,cm$^{-2}$ -- approximately two orders of magnitude higher than we find, as well as greater than the upper limit attained from the \textit{BeppoSAX} satellite mentioned in Section \ref{sec:target_selection}.

We further find this result to be inconsistent with \textit{NuSTAR}.  Using a simple \textsc{powerlaw\,+\,gaussian} model fitted to the \textit{NuSTAR} data, we obtain $F\,^{\textrm{FPMA}}_{15-50\,\textrm{keV}}=3.0\,^{+0.9}_{-0.5}\,\times\,10^{-12}$\,erg\,s$^{-1}$\,cm$^{-2}$ and $F\,^{\textrm{FPMB}}_{15-50\,\textrm{keV}}=3.1\,^{+1.0}_{-0.4}\,\times\,10^{-12}$\,erg\,s$^{-1}$\,cm$^{-2}$.  Extrapolating these fluxes to the 20\,--\,100\,keV band gives $F\,_{20-100\,\textrm{keV}}\,\sim\,1.8\,\times\,10^{-12}$\,erg\,s$^{-1}$\,cm$^{-2}$ for both Focal Plane Modules, which is fully consistent with the upper limit found with the \textit{BeppoSAX} satellite ($F^{BeppoSAX}_{\textnormal{20\,--\,100\,keV}}\leqslant$\,9.12\,$\times$\,10$^{-12}$\,erg\,s$^{-1}$\,cm$^{-2}$).

Figure \ref{fig:f2} shows the \textit{Suzaku} XIS spectra over-plotted with the \textit{NuSTAR} data.  The spectral data for \textit{Suzaku} XIS and \textit{NuSTAR} are consistent with each other in shape within the common energy range 3\,--\,10\,keV.  The composite spectrum formed spans approximately two dex in energy, from 0.7\,--\,34\,keV, as a result of the minimum SNR grouping procedures for each data set.

\subsection{\textit{Chandra}}
\label{sec:obs_CHA}

The \textit{Chandra} level 2 event file was obtained from the \textsc{heasarc} database.  A fraction of the total collecting area of the detector was used in the \textit{timed exposure mode} setting, where the CCD collects data for a set frame time.  Selecting a frame time less than the default value (3.2s for \textit{Chandra}) reduces the probability of \textit{pile-up}.  This is where more than one photon in a particular bin from the same event is detected.  An exposure time of 0.4\,s per frame was used in the observation of IC 3639, which gives a reduced predicted pile-up fraction of $\sim$\,0.3\%.  This setting was chosen in the original \textit{Chandra} observation proposal due to the previously unknown X-ray flux of the source, to minimise the risk associated with pile-up.

Spectral extraction from the \textit{Chandra} data was carried out with the \textsc{ciao 4.7} software package\footnote{http://cxc.harvard.edu/ciao/}.  We primarily investigated the \textit{Chandra} image of IC 3639 for potential contaminants located within the \textit{Suzaku} and \textit{NuSTAR} extraction regions.  However, no particularly prominent contaminating sources were visible in the immediate vicinity of the AGN.
A comparatively large circular source region of 2\farcm6 was used with the XIS image due to its larger point spread function (PSF) relative to \textit{Chandra}.  As a result, the XIS spectra will contain some flux from non-AGN related activity, unresolvable by that instrument.
To account for this, the \textit{Chandra} image was used to model as much of the unresolved non-AGN activity from the \textit{Suzaku} XIS image as possible.  An annular \textit{Chandra} source region with outer radius as close to that of the circular XIS source region as possible was created with inner region of radius 0\farcm05,  excluding the central AGN.  A simple power law was fitted to this spectrum, and was added to the model used with the \textit{Suzaku} XIS data.  This power law is referred to as the \textit{contamination power law (CPL)} hereafter.

Ideally, the outer radius of the annular extraction radius used in the \textit{Chandra} image would equal the radius of the circular source region used in the XIS image.  However, as noted above, data were taken with a custom 1\,/\,8 sub-array on ACIS-S3.  This meant that \textit{Chandra} only observed part of the sky covered by the other instruments.  As such, the image produced could not have a source region wider than $\sim$0\farcm5, as opposed to the XIS source region of radius 2\farcm6.
Accordingly, we used an annular region of outer radius 0\farcm5 for the \textit{Chandra} image. The annular counts extraction region and circular background region of equal radius used with the \textit{Chandra} image are shown in the right panel of Figure \ref{fig:1b}.  The $\tt specextract$ command was used to create the spectral and response files for use in \textsc{xspec}.  The \textit{Chandra} spectrum was grouped with greater than or equal to 20 counts per bin prior to the \textsc{cpl} modelling.  Furthermore, the flux of the CPL in the \textit{Chandra} energy band (0.5\,--\,8\,keV) was $F\,^{\textrm{CPL}}_{0.5-8.0\,\textrm{keV}}\,=\,7.17\,\times\,10^{-14}$\,erg\,s$^{-1}$\,cm$^{-2}$.

\section{X-RAY SPECTRAL FITTING}
\label{sec:x-ray_fitting}

The resulting spectral data sets for \textit{Suzaku} XIS and \textit{NuSTAR} were used in the energy ranges 0.7\,--\,9.0\,keV and 3.0\,--\,34.0\,keV, respectively.  Data above this threshold were excluded due to low SNR.  Initially, the \textit{Suzaku} XIS and \textit{NuSTAR} data were fitted independently of each other with a simple \textsc{powerlaw\,+\,gaussian} model to give the following fluxes in the 2\,--\,10\,keV energy band:\\

\textit{NuSTAR}:\\
$F\,^{\textrm{FPMA}}_{2\,-\,10\,\textrm{keV}}=1.81\,^{+0.16}_{-0.41}\,\times\,10^{-13}$\,erg\,s$^{-1}$\,cm$^{-2}$,\\
$F\,^{\textrm{FPMB}}_{2\,-\,10\,\textrm{keV}}=1.86\,^{+0.17}_{-0.40}\,\times\,10^{-13}$\,erg\,s$^{-1}$\,cm$^{-2}$.\\

\textit{Suzaku} XIS:\\
$F\,^{\textrm{XIS03}}_{2\,-\,10\,\textrm{keV}}=1.84\,\pm\,0.24\,\times\,10^{-13}$\,erg\,s$^{-1}$\,cm$^{-2}$,\\
$F\,^{\textrm{XIS1}}_{2\,-\,10\,\textrm{keV}}=2.12\,^{+0.34}_{-0.32}\,\times\,10^{-13}$\,erg\,s$^{-1}$\,cm$^{-2}$.\\

The overall match between these fluxes implies we can analyse the data sets together.  A \textit{Chandra} spectrum was extracted from a 3\farcs9 circular extraction region.  However, there were only 35 counts present in the 2\,--\,10\,keV band.  By using the same \textsc{powerlaw\,+\,gaussian} model to determine the flux as with the other data sets, we get:\\
$F\,^{\textrm{\textit{Chandra}}}_{2\,-\,10\,\textrm{keV}}=9.51\,^{+7.83}_{-9.51}\,\times\,10^{-14}$\,erg\,s$^{-1}$\,cm$^{-2}$.\\
This flux is only mildly inconsistent with the other data sets, but also consistent with zero, due to the low SNR of the data.  Given this low SNR, we do not use the \textit{Chandra} AGN data for further spectral analysis, but the consistency found between this and \textit{Suzaku}/\textit{NuSTAR} data sets suggests that we are classifying IC 3639 as a bona fide CTK AGN robustly.

The power-law slope of the composite spectrum (\textit{Suzaku}\,+\,\textit{NuSTAR}) is hard, with photon index $\mathrm{\Gamma}\sim1.8$ and the EW of the Fe-K$\alpha$ line is very large (EW\,$\sim 2.4$\,keV).  These are consistent characteristics of a heavily obscured AGN.  Due to the high EWs found for the Fe-K$\alpha$ line with a power-law\,+\,Gaussian model, we proceeded to fit more physically motivated models as follows.

A general model \textit{structure} was used with each spectrum, included in Equation \ref{eq:template}.  However, not all models used in this study required all the components listed in the template.
\noindent
\begin{gather}
    \textsc{\textbf{Template} = const $\times$ phabs[gal] $\times$ [apec + cpl + spl + \nonumber}\\
    \textsc{(refl + obsc $\times$ ipl + f\_lines)]}
	\label{eq:template}
\end{gather}
Below, we give explicit details for each component in Equation \ref{eq:template}:
\begin{itemize}
\item \textsc{const}: multiplying constant used to determine the cross-calibration between different instruments.  The \textit{NuSTAR} FPMA constant was frozen to unity and the other three constants left free (\citealt{Madsen2015} report cross-normalisation constants within 10\% of FPMA).

\item \textsc{phabs[gal]}: component used to account for photoelectric absorption through the Milky Way, based on $H_{\textrm I}$ measurements along the LOS \citep{Dickey1990}.  This is represented as an obscuring column density in units of cm$^{-2}$, and assumed constant between instruments (and so was frozen for each data set).  The determined value was 5.86\,$\times\,10^{20}\,\textnormal{cm}^{-2}$.

\item \textsc{apec} \citep{Smith2001}: model component used as a simple parameterisation of the softer energy X-ray emission associated with a thermally excited diffuse gas surrounding the AGN.  Detailed studies of brighter local AGN indicate that photoionisation may provide a better description of the soft X-ray emission in AGN spectra (e.g. \citealt{Guainazzi2009,Bianchi2010}), but such modelling would require a higher SNR and better spectral resolution than currently available for IC 3639.  The low-energy spectral shape for IC 3639 found with \textit{Suzaku} XIS is far softer than for the higher energy portion of the spectrum.

\item \textsc{cpl}: component referring to the \textit{contamination power law}, used to account for the unresolved non-AGN emission \textit{contaminating} the \textit{Suzaku} XIS spectral counts.  See Section \ref{sec:obs_CHA} for further details.

\item \textsc{spl}: component referring to the \textit{scattered power law}.  This accounts for \textit{intrinsic} AGN emission that has been scattered into our LOS from regions closer to the AGN, such as the NLR.  The power law photon index and normalisation were tied to the intrinsic AGN emission as a simplification.  However, a constant multiplying the \textsc{spl} component was left free to allow a variable fraction of observed flux arising from scattered emission.

\item The final term in Equation \ref{eq:template} collectively consists of three parts:  

\begin{itemize}
\item \textsc{refl}: reflected component, arising from the primary nuclear obscurer and has been modelled in varying ways.  In this work, we use slab models (\textsc{pexrav} and \textsc{pexmon}) as well as toroidal geometry models (\textsc{torus} and \textsc{mytorus}), described in Sections \ref{sec:slabs}, \ref{sec:T} and \ref{sec:M} respectively.
\item \textsc{obsc $\times$ ipl}: Most models include the direct transmitted component (\textit{intrinsic power law} or \textsc{ipl}), after accounting for depletion due to absorption through the obscurer via the multiplying \textsc{obsc} term.
\item \textsc{f\_lines}: component describing fluorescence lines believed to arise from photon interactions with the circumnuclear obscurer.
\end{itemize}
\end{itemize}

\subsection{Slab models: \textsc{pexrav} and \textsc{pexmon}}
\label{sec:slabs}

Slab models describe X-ray reflection off an infinitely thick and long flat slab, from a central illuminating source.  \textsc{pexrav} \citep{Magdziarz1995} comprises an exponentially cut-off power law illuminating spectrum reflected from neutral material.  To acquire the reflection component alone, with no direct transmitted component, the reflection scaling factor parameter is set to a value $R$\,\textless\,0.
Other parameters of interest include the power law photon index; cutoff energy; abundance of elements heavier than helium; iron abundance (relative to the previous abundance) and inclination angle of the slab (90$^{\circ}$ describes an edge-on configuration; 0$^{\circ}$ describes face-on).  The model gave far better reduced chi-squared values for a reflection-dominated configuration, and as such the reflection scaling factor was frozen to $-1.0$, corresponding to a 50\% covering factor.
\textsc{pexrav} does not self-consistently include fluorescent line emission.  As such, a basic Gaussian component was initially added to account for the strong Fe-K$\alpha$ line, resulting in an EW\,$\sim1.4-3.0$\,keV.  Alternatively, the \textsc{pexmon} model \citep{Nandra2007} combines \textsc{pexrav} with approximated fluorescence lines and an Fe-K$\alpha$ Compton shoulder \citep{Yaqoob2011}.  The fluorescence lines include Fe-K$\alpha$, Fe-K$\beta$ and nickel-K$\alpha$.  All analysis with slab models refer to the \textsc{pexmon} model hereafter, and are denoted as model \textbf{P}.  The high EWs found for the Fe-K$\alpha$ line with the reflection-dominated \textsc{pexrav}\,+\,Gaussian model in addition to the power-law\,+\,Gaussian model, we next considered more physically motivated self-consistent obscured AGN models.

\subsection{\textsc{bntorus}}
\label{sec:T}

Two tabular models are provided by \citet{Brightman2011} to describe the obscurer self-consistently including the intrinsic emission and reflected line components.  The spherical version describes a covering fraction of one in a geometry completely enclosing the source.  The presence and morphology of NLRs in a multitude of sources favours a covering factor \textless\,1, implying an anisotropic geometry for the shape of the obscurer in most Seyfert galaxies.  For this reason, preliminary results were developed with this model, before analysis was carried out with toroidal models.  For further discussion of the NLR of IC 3639, see Section \ref{sec:dis_spec_comps}.

The second \textsc{bntorus} model (model \textbf{T} hereafter) was used extensively in this study.  This models a toroidal obscurer surrounding the source, with varying opening and inclination angles.  Here, the opening angle describes the conical segment extending from both poles of the source (i.e. the half-opening angle).  Because the obscurer is a spherical section in this model, the column density along different inclination angles does not vary.  The range of opening angles studied is restricted by the inclination angle since for inclination angles less than the opening angle, the source becomes unobscured.  Thus to allow exploration of the full range of opening angles, we fixed the inclination angle to the upper limit allowed by the model: 87$^{\circ}$ \citep{Brightman2015}.  The tables provided for \textsc{bntorus} are valid in the energy range 0.1\,--\,320\,keV, up to obscuring column densities of $10^{26}$\,cm$^{-2}$.  Equation \ref{eq:T} describes the form of model T used in \textsc{xspec} - all properties associated with the absorber are present in the \textsc{torus} term, and collectively used in the modelling process.

\begin{gather}
    \textsc{model \textbf{T} = const $\times$ phabs $\times$ (apec + spl + \nonumber}\\
    \textsc{cpl + torus)}
	\label{eq:T}
\end{gather}

\citet{Liu2015} report model T to over-predict the reflection component for edge-on geometries, resulting in uncertainties.  However, varying the inclination angle did not drastically alter our fits and consistent results were acquired between both models T and M, described next.

\subsection{\textsc{mytorus}}
\label{sec:M}

The \textsc{mytorus} model, developed by \citet{Murphy2009}, describes a toroidal-shaped obscurer with a fixed half-opening angle of 60$^{\circ}$, and free inclination angle.  However, because the geometry here is a doughnut as opposed to a sphere, the LOS column density will always be less than or equal to the equatorial column density (with equality representing an edge-on orientation with respect to the observer).

The full computational form of this model is shown in Equation \ref{eq:M}, and encompasses the energy range 0.5\,--\,500\,keV, for column densities up to $10^{25}$\,cm$^{-2}$.  Three separate tables are used to describe this model: the \textit{transmitted absorption} or \textit{zeroth order} continuum, altered by photoelectric absorption and Compton scattering; the \textit{scattered component}, describing Compton scattering off the torus; and the \textit{fluorescence emission} for neutral Fe-K$\alpha$ and Fe-K$\beta$ together with their associated Compton shoulders.  This study uses the \textit{coupled} mode for this model, where model parameters are tied between different table components (referred to as model \textbf{M} hereafter).  For further details on the decoupled mode, which is often used for sources showing variability or non-toroidal geometries with high SNR data, refer to the publicly available \textsc{mytorus} examples,\footnote{http://mytorus.com/mytorus-examples.html} or see \citet{Yaqoob2012}.

\begin{gather}
    \textsc{model \textbf{M} = const $\times$ phabs $\times$ (apec + spl + cpl + \nonumber}\\
    \textsc{pow * etable \{\tt trans. absorption\} + \nonumber}\\
    \textsc{atable\{\tt scattered\} + \nonumber}\\
    \textsc{atable\{\tt fluor\_lines\})}
	\label{eq:M}
\end{gather}

\section{Results from spectral fitting}
\label{sec:results}

In this section, we present the results of our X-ray spectral fitting of IC 3639 together with model-specific parameters shown in Table \ref{tab:obs_parameters}.  Figures \ref{fig:f4a} and \ref{fig:f4b} show the spectra and best-fit models attained for models T and M, respectively.
First we consider the EW of the Fe-K$\alpha$ line. As previously mentioned, an EW of the order of 1\,keV can be indicative of strong reflection.  \citet{Risaliti1999} found the EW for IC 3639 to be $3.20^{+0.98}_{-1.74}$\,keV.  In order to determine an EW for the Fe-K$\alpha$ line here, we modelled a restricted energy range of $\sim$3\,--\,9\,keV with a simple \textsc{(powerlaw\,+\,gaussian)} model.  Here the power law was used to represent the underlying continuum, and the Gaussian was used as a simple approximation to the Fe-K$\alpha$ fluorescence line.  Additionally, all four data sets were pre-multiplied by cross-calibration constants in the same way as described in the template model.

Due to low signal-to-noise, the continuum normalisation had a large uncertainty.  As such, a robust error could not be directly determined on the EW using \textsc{xspec}.  Alternatively, we carried out a four dimensional grid to step over all parameters of the model in \textsc{xspec}, excluding line energy, which was well defined and frozen at 6.36\,keV in the observed frame.  The EW was calculated for each grid value, and the corresponding confidence plot is shown in Figure \ref{fig:f5a}.  The horizontal black line represents the 90\% confidence region for the chi-squared difference from the best-fit value, $\Delta\chi^2$.  Here, the 90\% confidence level refers to the chi-squared distribution for four free parameters with value $\Delta\chi^2$\,=\,7.779.  Figure \ref{fig:f5b} shows the model used, fitted to the four data sets.  This gave an EW of $2.94^{+2.79}_{-1.30}$\,keV, consistent with \citet{Risaliti1999}.  This is well above the approximate threshold of 1\,keV typically associated with the presence of CTK obscuration.  However, \citet{Gohil2015} find the presence of dust in the obscuring medium can enhance the Fe-K$\alpha$ line detection even for CTN gas.  This is a further reason for the importance of consistent modelling to determine the column density more robustly.  Additionally, the errors seem to favour a high EW, with the upper limit fully encapsulating the most extreme cases reported by \citet{Levenson2002}.

\newpage
\begin{subfigures}
\begin{figure*}
\centering
  \includegraphics[angle=-90,width=0.8\textwidth]{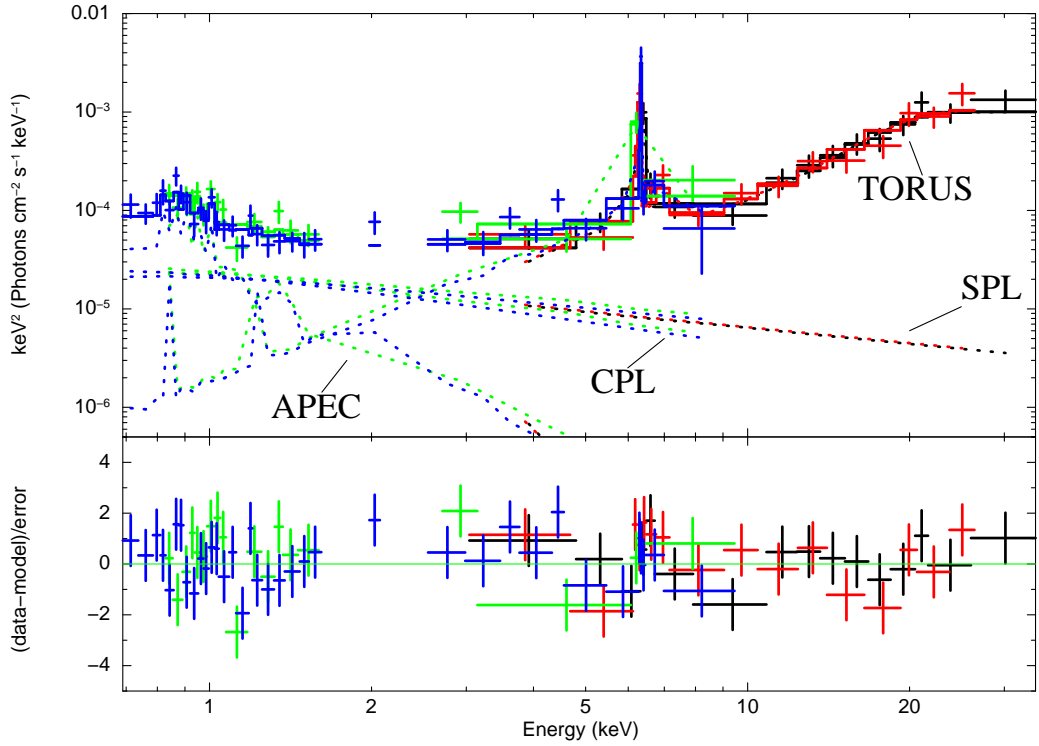}
  \caption{\label{fig:f4a}Best fit for spectral model T.  Top panel represents the unfolded spectrum, with the bottom panel representing the residuals present.  Both fits appear to show a slight residual still present around the iron line energy band of $\sim$6.4\,keV.  Model components are labelled in accordance with the description in Section \ref{sec:T}.  Colour scheme is the same as in Figure \ref{fig:f2}.}
\end{figure*}
\begin{figure*}
\centering
  \includegraphics[angle=-90,width=0.8\textwidth]{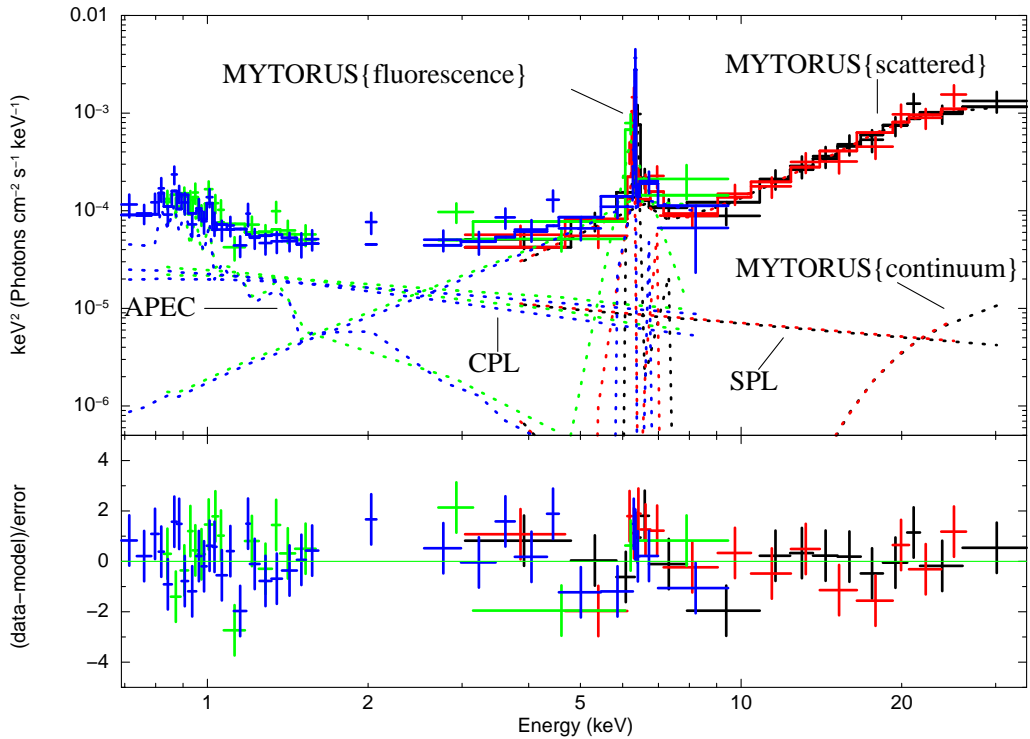}
  \caption{\label{fig:f4b}Best fit for spectral model M.  Layout and colour scheme is the same as in Figure \ref{fig:f4a}.Top panels in both represent the unfolded spectrum, with the bottom panel representing the residuals present.  The fit again appears to show a slight residual present around the iron line energy band of $\sim$6.4\,keV.  Model components are labelled in accordance with the description in Section \ref{sec:M}.  Although the components are labelled separately, all parameters between the tables for these components were tied together in the default \textit{coupled mode} for the model - see the text for further details.}
\end{figure*}
\label{fig:f4}
\end{subfigures}

\newpage
\begin{subfigures}
\begin{figure*}
\centering
  \includegraphics[width=0.7\textwidth,trim={1cm -1cm 1cm 1cm}]{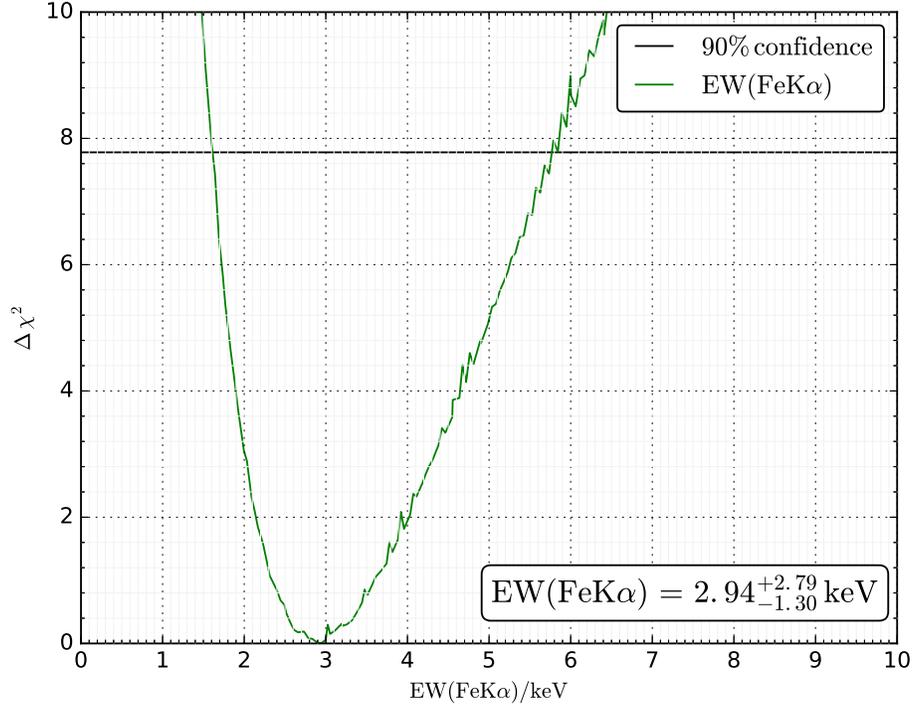}
  \caption{\label{fig:f5a}EW of the Fe-K$\alpha$ line as a function of $\Delta \chi^2$.  The horizontal black line represents the 90\% confidence level for the chi-squared distribution with four free parameters, of $\Delta \chi^2$\,=\,7.779.  This was generated with a four dimensional grid over the best-fit parameters used in the model.}
\end{figure*}
\begin{figure*}
\centering
  \includegraphics[trim={0cm 0cm 0cm 0cm},angle=-90,width=0.7\textwidth]{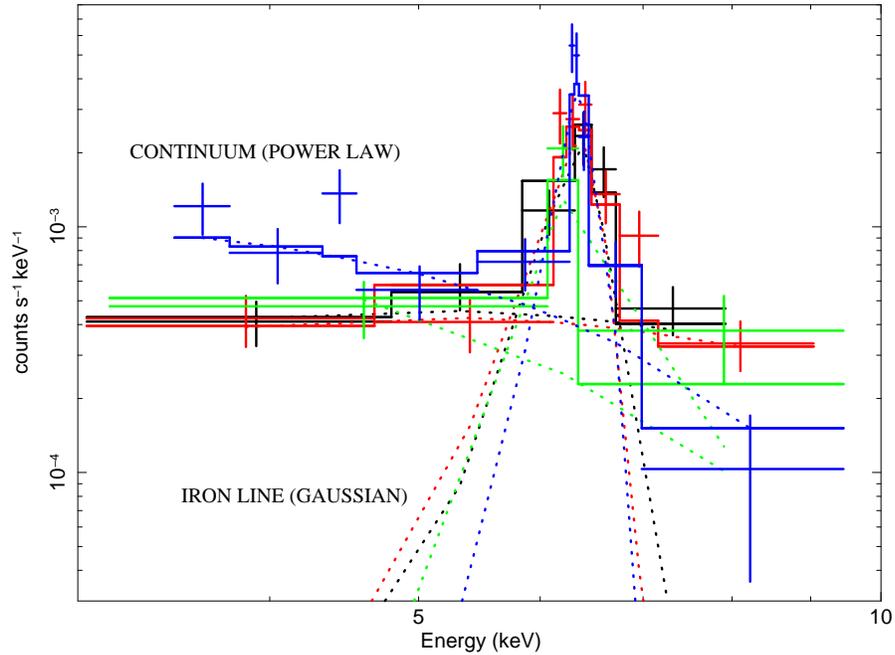}
  \caption{\label{fig:f5b}Fit generated with the simplified model for all four data sets, including cross-calibration constants.  The colour scheme used is the same as in Figure \ref{fig:f2}.  Dotted lines show individual additive model components used, labelled for clarity.}
\end{figure*}
\label{fig:f5}
\end{subfigures}

Both models T and M yield consistent cross-calibration constants between data sets (see Table \ref{tab:obs_parameters}), with the exception of cross-calibration between \textit{NuSTAR} FPMA and \textit{Suzaku} XIS1.  The cross-calibration constant between \textit{NuSTAR} and \textit{Suzaku} data significantly deviated from unity if the \textsc{cpl} component were removed.  This strongly indicates that the extra \textsc{cpl} component is necessary.  The varying cross-calibration between \textit{Suzaku} and \textit{NuSTAR} may be due to instrumental differences unaccounted for with the \textsc{cpl} component, or perhaps a subtle signature of variability.  To test these results, the cross-calibration constants were fixed to similar results found in \citet{Madsen2015}.  This resulted in comparable reduced chi-squared values of 98\,/\,80 and 104\,/\,80 for models T and M respectively, together with marginally altered physical parameters from those presented in Table \ref{tab:obs_parameters}.

The soft emission was modelled with \textsc{apec}.  The values of \textit{kT} found for either model show strong agreement, both being consistent with 0.8\,keV within errors.  Varying other parameters did not significantly alter this value or its corresponding normalisation.  Agreement for the \textsc{apec} component between models T and M is expected from Figures \ref{fig:f4a} and \ref{fig:f4b}, since this dominates the other model components at soft enough energies for both models.  The corresponding intrinsic soft luminosity (solely from the \textsc{apec} component) in the 0.5\,--\,2\,keV band was consistently found to be $2.0\,\times\,10^{40}$\,erg\,s$^{-1}$.  Note the \textsc{apec} flux is $\sim$3 times the flux derived from the \textsc{cpl} component.

The scattering fraction (numerically represented by the constant multiplying the \textsc{spl} component) is comparable between models T and M.  Even within the high upper limit found for either model, the total scattering fraction is $\lesssim$\,0.6\%.  Such values are not uncommon in previous CTK studies (e.g. \citealt{Gandhi2015,Annuar2015}), and suggest that a minor contribution of the total flux arises from scattered emission here, although proper modelling of higher SNR data describing the soft emission would be required to better constrain this.

Next we consider parameters relating to the absorber specifically.  The equatorial column density for model T is the same as the column density along the LOS, whereas the LOS column density for model M is less than or equal to the equatorial column density.  This is the reason for the two separate entries in Table \ref{tab:obs_parameters} for model M.  Model T indicates a strongly CTK obscuring column density of $9.0 \times 10^{\textrm{24}}\,\textrm{cm}^{\textrm{-2}}$ along the LOS.  For comparison, model M gives a similar LOS column density at $9.8 \times 10^{\textrm{24}}\,\textrm{cm}^{-2}$.  Both models are unconstrained at the upper limit and also \textgreater\,$3.0 \times 10^{\textrm{24}}\,\textrm{cm}^{\textrm{-2}}$ for the lower limit, consistently within the CTK regime (see Table \ref{tab:obs_parameters}).

Initially the inclination angle and opening angle were left free to vary in model T, but this led to the model diverging to the limits - i.e. the upper limit on inclination angle (describing an edge-on torus) and the lower limit on opening angle (describing a large covering fraction).  The inclination angle for both models was tested by stepping over the parameter in \textsc{xspec} in the full allowable range, in addition to fixing the angle to intermediate values such as 60$^{\circ}$.  This did not result in a significant improvement in $\Delta \chi^2$, and in some cases worsened the fit.  As discussed in Section \ref{sec:T}, the inclination angle of model T was fixed to 87$^{\circ}$ to allow exploration of a full range of opening angles.  In contrast, model M has a fixed half-opening angle (by default) and the inclination angle was left free. The inclination angle found for model M is lower than for model T, at $\sim$\,84$^{\circ}$, inconsistent with model T at the upper end.  This could be affected by the model inconsistencies at edge-on inclinations for model T reported by \citet{Liu2015}.  This still suggests a near edge-on torus inclination however.  In contrast, the opening angle for model T (29$^{\circ}$) is lower than the fixed value in model M.  A reduced opening angle implies an increased covering factor surrounding the source and thus potentially a strengthened reflection component.

The intrinsic AGN spectrum can be studied via the continuum photon index.  Both models consistently agree on a soft photon index of $\sim$2.5 - far softer than the average value of $\sim$1.9 found in large surveys (e.g. \citealt{Mateos2005}).  However, our value is consistent with typical values within the uncertainties.  To test this, the photon index was fixed to 1.9 in both models.  The $\Delta \chi^2$ values increased to 97\,/\,78 and 101\,/\,78, yielding F-test statistics of 3.05 and 1.72 for models T and M, respectively.  These values suggest that a photon index of 1.9 is marginally less likely, but not immediately ruled out in either case.  Such high photon indices have been found before from the torus models used with CTK sources \citep{Balokovic2014,Brightman2015} and could imply accretion at a large fraction of the Eddington rate \citep{Brightman2016}.  The Eddington ratio is discussed further in Section \ref{sec:dis_L}.  Additionally, the absorber is likely more complex in reality than a geometrically smooth torus as assumed in models T and M (coupled).  This has been found in NGC 1068, by \citet{Bauer2015} for example, where a multi-component reflector is comprised of several layers of differing column densities.  We include in the Appendix a contour plot between the intrinsic photon index and column density for models T and M as an example.  The plots both show the unconstrained nature of $N_{\textrm{H}}$ as well as the favoured soft photon index by either model.

Overall both models T and M give acceptable fit statistic $\chi^2\mathbin{/}\textrm{dof}$ values of 94\,/\,77 and 99\,/\,77, respectively.  Initial testing with model P yielded a lower value of reduced chi-squared of 85\,/\,76.  Since the transmitted power law is not directly visible over any of the spectrum, constraining the reflection fraction (defined as the strength of the reflection component relative to a semi-infinite slab subtending $2\pi$\,steradians on the sky, fully reflecting the intrinsic power law) is highly uncertain.  This was used as justification to fix the reflection scaling factor to -1.0.  Other than the reflection dominated nature of the source, there are few aspects to be learnt from the over simplified slab geometry of \textsc{pexmon}.  Furthermore, slab models effectively give a lower limit on the intrinsic power of the source, since the slab subtends 2$\pi$ steradians on the sky, equivalent to a 50\% unobscured covering factor, as opposed to the torus models, in which this solid angle is computed self-consistently with inclination.
Model P did, however, appear to require a super-Solar iron abundance to explain the prominent iron line complex present in the spectra of IC 3639.  The iron abundance (defined in units of the Solar abundance) and abundance of elements heavier than helium (defined in units of the iron abundance) were tied to each other and left free.  This yielded an abundance of $2.0^{+0.7}_{-0.5}$.  We tested this outcome by freezing the abundance and iron abundance to Solar values, as is default in models M and T.  This resulted in a considerable increase in reduced chi-squared to 102\,/\,78.  Fixing one of these individually of the other resulted in comparable best-fit reduced chi-squared values, but with the free parameter of the two significantly deviating from 1.0.  In comparison, the fits shown in Figures \ref{fig:f4a} and \ref{fig:f4b}, using the toroidal models T and M respectively, show a slight residual around the iron line region.  This suggests both models are insufficiently describing the iron fluorescence.  Besides strong reflection, high iron abundance is one possible cause of prominent iron fluorescence and may be partly responsible for the extreme Fe-K$\alpha$ line EW observed for IC 3639.  Alternatively, \citet{Levenson2002} discuss how circumnuclear starbursts can also lead to strong iron emission.  This is analysed further in Section \ref{sec:dis_spec_comps}, where the star formation rate (SFR) is considered.

\begin{table*}
	\begin{center}
 	\caption{Parameters determined from the X-ray spectral fitting of IC 3639.}
 	\begin{tabular}{rcccl}
  		\hline \hline
 		Component 								& Parameter 								& Model T 						& Model M 						&Units\\[0.1cm]
 		\hline 
 		Fe-K$\alpha$ fluorescence emission-line & Equivalent width 							&\multicolumn{2}{c}{..........$2.72^{+2.99}_{-0.93}$..........}	&keV\\[0.1cm]
 		\hline %
 		Cross-calibration constants 			& $\textrm{[FPMA$\mapsto$FPMB]}$	& $1.02^{+0.14}_{-0.15}$		& $1.02^{+0.15}_{-0.14}$		&-\\[0.1cm]
 												& \multicolumn{1}{l}{$\textrm{[FPMA$\mapsto$XIS1]}$}	& $1.21^{+0.25}_{-0.21}$		& $1.25^{+0.25}_{-0.23}$		&-\\[0.1cm]
 												& \multicolumn{1}{l}{$\textrm{[FPMA$\mapsto$XIS03]}$}	& $1.11^{+0.21}_{-0.16}$		& $1.14^{+0.20}_{-0.19}$		&-\\[0.1cm]
 		\hline %
 		Soft emission (\textsc{apec})			& \textit{kT} 										& $0.79^{+0.13}_{-0.09}$		& $0.78^{+0.08}_{-0.10}$		&keV\\[0.1cm]
 												& $L^{\textrm{int}}_{0.5-2\textrm{ keV}}$$^{\dagger}$		& $2.01$		& $2.06$		&$\times\, 10^{40}$erg s$^{-1}$\\[0.1cm]
 		\hline %
 		Diffuse scattering fraction (\textsc{spl})& $f_{\textrm{scatt}}$ 					& $0.97^{+3.39}_{-0.63}$		& $0.20^{+5.58}_{-0.15}$		&$\times\, 10^{-3}$\\[0.1cm]
 		\hline %
 		Column densities						& $N_{\textrm{H}}$(eq) 						& \multirow{2}{*}{$8.98^{+\textrm u}_{-3.23}$} 		& $10.0^{+\textrm u}_{-4.1}$							&\multirow{2}{*}{$\times\, 10^{24}$cm$^{-2}$}\\[0.1cm]
 												& $N_{\textrm{H}}$(los) 					& 		&$9.76^{+\textrm u}_{-6.15}$\\[0.1cm]
 		\hline
 		Orientation Angle 						& $\theta_{\textrm{inc}}$					& $87.0^{\textrm{f}}$		& $83.8^{+1.9}_{-17.2}$		&\multirow{2}{*}{deg}\\[0.1cm]
 		Half-opening Angle							& $\theta_{\textrm{tor}}$ 					& $28.5^{+26.1}_{-\textrm u}$							&$60.0^{\textrm{f}}$\\[0.1cm]
 		\hline %
 		AGN continuum 							& $\Gamma_{\textrm{int}}$					& $2.54^{+0.27}_{-0.33}$		& $2.46^{+u}_{-0.60}$		&-\\[0.1cm]
 												& $L^{\textrm{int}}_{2-10\textrm{ keV}}$$^{\dagger}$ 								& $9.26$		& $45.7$		&$\times\, 10^{42}$erg s$^{-1}$\\[0.1cm]
 												& $L^{\textrm{int}}_{0.5-30\textrm{ keV}}$$^{\dagger}$ 								& $2.99$		& $14.0$		&$\times\, 10^{43}$erg s$^{-1}$\\[0.1cm]
 		\hline %
 		$\chi^2\mathbin{/}\textrm{dof}$			&											& $94/77$					& $99/77$					&-\\[0.1cm]
 		\hline 
 	\multicolumn{5}{p{.8\textwidth}}{$^{\textrm{f}}$ - fixed values.}\\
 	\multicolumn{5}{p{.8\textwidth}}{$\Gamma_{\textrm{int}}$ - intrinsic AGN photon index, determined for the range 1.4\,--\,2.6.}\\
 	\multicolumn{5}{p{.8\textwidth}}{$^{\dagger}$luminosities calculated with a distance to the source of 53.6 Mpc, from the best-fit values determined in \textsc{xspec}.}\\
 	\multicolumn{5}{p{.8\textwidth}}{u - unconstrained.}\\
 	\end{tabular}
 	\end{center}
 	\label{tab:obs_parameters}
 \end{table*}

\section{DISCUSSION}
\label{sec:discussion}

\subsection{Spectral components}
\label{sec:dis_spec_comps}
The LOS obscuring column densities for models M and T are consistent with one another, both well within the CTK regime and unconstrained at the upper limit.  Our findings are also consistent with \citet{Risaliti1999}, arguing against source variability between the \textit{NuSTAR} and \textit{Suzaku} observations.

The column density determined here establishes IC 3639 as a CTK AGN in a face-on host-galaxy.  Such a configuration is uncommon but not unheard of (e.g. \citealt{Annuar2015}).  However, \citet{Fischer2013} find no correlation between the orientations of the NLR and host-galaxy disc suggesting that the obscurer thought to be responsible for shaping the NLR in many galaxies may be independent of the host disk.  Furthermore, \citeauthor{Fischer2013} find IC 3639 to have ambiguous NLR kinematics.  This is where targets display a symmetrical ionised gas component on either side of the nucleus, but uncertainty remains as to whether or not these represent each half of a NLR bicone.  A non-biconical outflow is consistent with heavy obscuration and could indicate a high covering factor, restricting NLR emission.

\citet{Levenson2002} find the highest Fe-K$\alpha$ EWs for sources with $N_\textrm{H}\sim6\times10^{24}\,\textrm{cm}^{-2}$, in combination with large inclination angles.  However, from simulations, the authors found that the EW diminishes at even higher column densities (since the fluorescence photons cannot escape for such high optical depths), comparable with the $N_\textrm{H}$ values determined here.  It should be noted that their simulations are for a more simplistic geometry formulation with a square torus cross section (although \citealt{Yaqoob2010} found CTK lines-of-sight gave Fe-K$\alpha$ line strengths considerably less than the maximum possible for a given geometry).  So this may indicate a secondary source of strong iron fluorescence for IC 3639, such as super-Solar iron abundance.  As already stated, this was found in model P to help fit the residuals present in the iron-line energy region.  Since models T and M assume Solar abundance, the final residuals present in Figures \ref{fig:f4a} and \ref{fig:f4b} around the iron line complex may be due to a super-Solar iron abundance present in IC 3639, or perhaps a high SFR.  Such high elemental abundances have been postulated to arise from different astrophysical events, such as high supernovae type Ia rates in the host-galaxy.  Alternatively, \citet{Ricci2014} find that as the column density for CTK AGN is increased, the EW of the iron fluorescence line is decreased.  This indicates a suppression of the reflection component for \textit{heavily} obscured systems, and suggests that the intrinsic iron line EW of IC 3639 could be even greater than we are observing here.

Regarding SFR, the intrinsic soft band (0.5\,--\,2\,keV) luminosity found from the \textsc{apec} component was $\sim2.0\,\times\,10^{40}\,\textnormal{erg\,s}^{\textnormal{-1}}$ for both models.  \citet{Mineo2012} detail a conversion between soft X-ray luminosity and host-galaxy SFR.  The authors determined the soft X-ray luminosity through the \textsc{mekal} model component, but for our purposes using the \textsc{apec}-determined luminosity is sufficient to establish an order of magnitude estimate.  By accounting for the dispersion in the \citet{Mineo2012} relation of 0.34 dex, we find $\textrm{SFR}_{\textrm{X-ray}}\,=\,39^{+46}_{-21}\,\textrm{M}_{\odot}\,\textrm{yr}^{-1}$.

Using a total IR luminosity calculated from the \textit{IRAS} catalogued fluxes\footnote{http://irsa.ipac.caltech.edu/applications/Gator/index.html} of $L_{8-1000\,\mu m}\,=\,8.14 \times 10^{\textrm{10}}\,\textrm{L}_{\odot}$, we find an IR-derived SFR using the relation presented by \citet{Murphy2011} to be $\textrm{SFR}_{\textrm{IR}}$\,$\sim$\,$12\,\textrm{M}_{\odot}\,\textrm{yr}^{-1}$.

Alternatively, polycyclic aromatic hydrocarbon (PAH) features are believed to be prominent in the spectra of starburst galaxies, with a good correlation found between PAH strength and IR luminosity.  We use Equation (5) from \citet{Farrah2007} to calculate a PAH-derived SFR (this equation uses an approximate scaling to account for the high rate of star formation  observed in ultraluminous infrared galaxies).  Using the 6.2 and 11.2\,$\mu$m luminosities for IC 3639 presented in \citet{Wu2009} (based on \textit{Spitzer}/IRS data).  We find an IR(PAH)-derived SFR of $\textrm{SFR}_{\textrm{IR(PAH)}}\,=\,12\,\pm\,6\,\textrm{M}_{\odot}\,\textrm{yr}^{-1}$, fully consistent with the \textit{IRAS}-derived value.

The X-ray SFR is higher than the IR(PAH) and IR(IRAS) SFRs by a factor of $\sim$\,2\,--\,3, but fully consistent within the uncertainties.  All SFRs determined here for IC 3639 are comparable with typical starburst galaxy SFRs determined and studied by \citet{Brandl2006}.  Furthermore, although \citet{Barnes2001} find the interacting galaxy group hosting IC 3639 to be free of a strong merger, they still report the possibility of enhancing star formation via galaxy harassment.  All of these factors are consistent with the hypothesis of \citealt{Levenson2002} that circumnuclear starbursts may lead to strong iron emission.

\subsection{Intrinsic AGN luminosity}
\label{sec:dis_L}

\begin{figure*}
\centering
\includegraphics[width=0.95\textwidth]{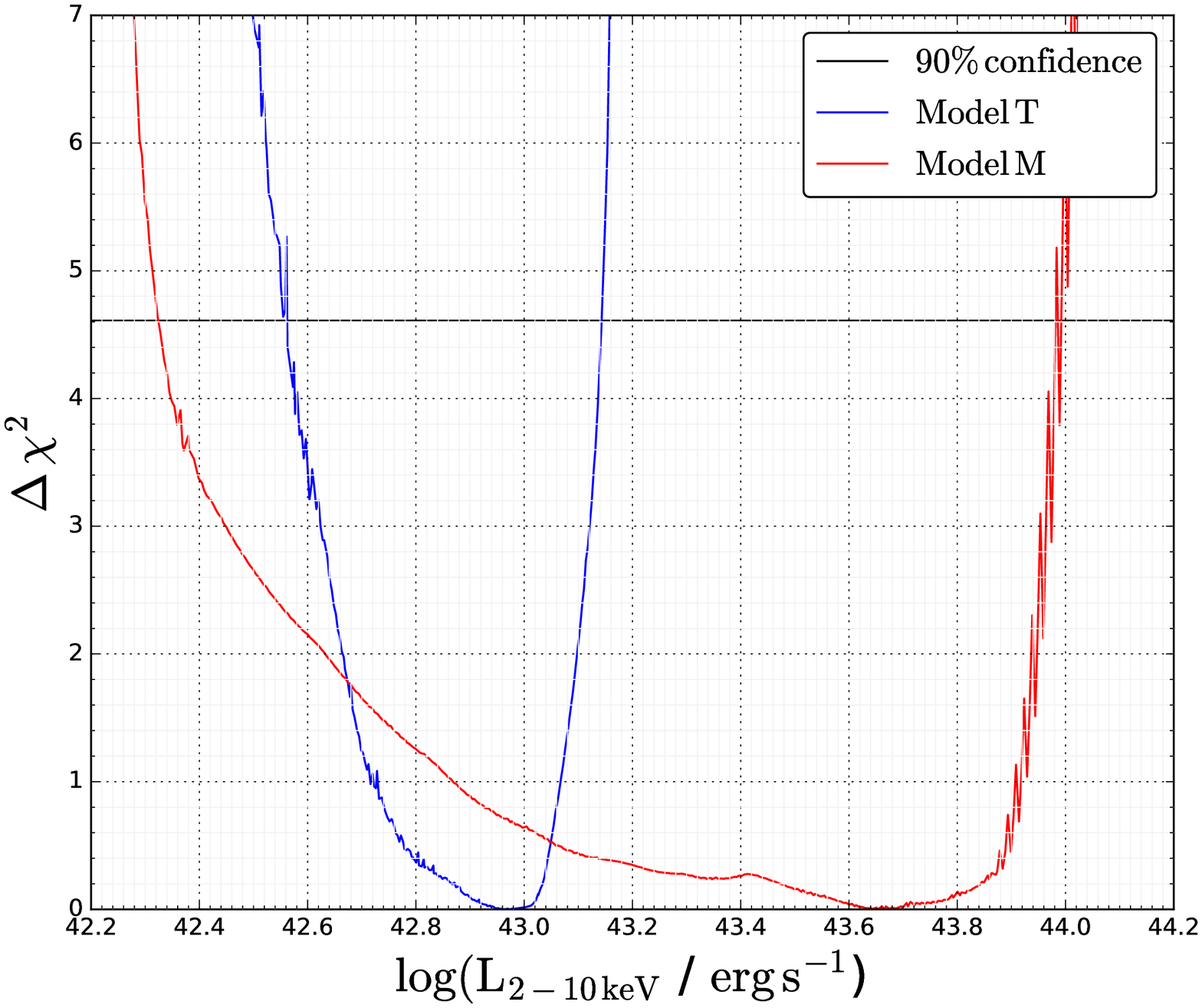}
\caption{Intrinsic X-ray luminosity (2\,--\,10\,keV) against $\Delta\chi^2$, corresponding to the difference between observed chi-squared value for a particular implementation of parameters and the best-fit chi-squared value corresponding to the best-fit parameter values presented in Table \ref{tab:obs_parameters}.  The 90\% confidence level is shown as a black line at $\Delta\chi^2$\,=\,4.61, in correspondence with the chi-squared distribution for two free parameters.  Model M encompasses the full range of model T luminosities found.}
\label{fig:f6}
\end{figure*}

The unobscured luminosity of the source in the 2\,--\,10\,keV band (the \textit{intrinsic} emission) was calculated with the model-dependent photon index and normalisation of the \textsc{ipl} component.  By stepping over the photon index and normalisation for either model in a two-dimensional grid, the intrinsic X-ray luminosity and corresponding $\Delta\chi^2$ value were determined.  The \textit{envelope} of all $\Delta\chi^2$ values for any given luminosity was then extracted, and is plotted in Figure \ref{fig:f6} for models T and M, similar to the four-dimensional grid used in Figure \ref{fig:f5a} to determine the EW of the iron line.  This gives a luminosity range of $\textrm{log}_{10}(L_{\textrm{2\,--\,10\,keV}}\,\textrm{[erg\,s}^{-1}])\,=\,42.3$\,--\,$44.0$, with intrinsic average X-ray luminosity $\textrm{log}_{10}(L_{\textrm{2\,--\,10\,keV}}\,\textrm{[erg\,s}^{-1}])\,=\,43.4^{+0.6}_{-1.1}$.  As can be seen in Figure \ref{fig:f6}, luminosities for model M completely encompass luminosities for model T at 90\% confidence.  Additional tests appear to show that this wide range of allowable model M luminosities is due to an uncertain inclination angle.  For example, we fixed the inclination angle to intermediate values in the range 70\,--\,84$^{\circ}$ for model M (approximate lower and upper limits found for the best fit).  The envelope presented in Figure \ref{fig:f6} fully encompassed the intermediate fixed inclination angle results.
Furthermore, a three-dimensional parameter space analysis between inclination angle, photon index and normalisation showed an increase of intrinsic X-ray luminosity with inclination angle, but also an increase in best-fit chi-squared value.

Recent works have demonstrated a correlation between X-ray luminosity and accretion disc luminosity.  For example, we use Equation (6) from \citet{Marchese2012} to approximate the accretion disc luminosity of IC 3639.  In order to consistently use this relation which is calculated to the 1\,--\,$\sigma$ confidence level, we have derived the 2\,--\,10\,keV luminosity for IC 3639, based on Figure \ref{fig:f6}, but at the 1\,--\,$\sigma$ confidence level for the chi-squared distribution with two free parameters of $\Delta\chi^2$\,=\,2.30.  This gives $\textrm{log}_{10}(L_{\textrm{2-10\,keV}}\,\textrm{[erg\,s}^{-1}])\,=\,43.4^{+0.6\,(1\sigma)}_{-0.8\,(1\sigma)}$, resulting in a disc luminosity:\\

$\textrm{log}_{10}(L_{\textrm{disc}}\,\textrm{[erg\,s}^{-1}])\,=\,44.5^{+0.7(+0.1)}_{-0.9(-0.2)}$.\\

\noindent The upper and lower bounds in brackets represent the intrinsic scatter from the \citeauthor{Marchese2012} relation, based on treating $L_{\textnormal{disc}}$ or $L_{\textnormal{2\,--\,10\,keV}}$ as the independent variable.  The other uncertainty represents the error associated with the observed 2\,--\,10\,keV luminosity uncertainty.

To determine the black hole mass ($M_\textrm{BH}$), we used the stellar velocity dispersion from \citet{Marinucci2012} of $99\,\pm\,5\,\textnormal{km}\,\textnormal{s}^{-1}$ with the M\,--\,$\sigma$ relation from \citet{Gultekin2009} to give $\textrm{log}_{10}(M_{\textrm{BH}} [M_{\odot}])=6.8\,\pm\,0.2$, and thus $\textrm{log}_{10}(L_{\textrm{Edd}}\,\textrm{[erg\,s}^{-1}])\,=\,44.9\,\pm\,0.2$.  This corresponds to an Eddington ratio of:\\

$\textrm{log}_{10}(\lambda_{\textrm{Edd}})=-0.4^{+0.8}_{-1.1}$,\\

\noindent to the 1\,--\,$\sigma$ confidence level.  Here we have defined $\textrm{log}_{10}(\lambda_{\textrm{Edd}})=\textrm{log}_{10}\Bigg(\displaystyle \frac{L_{\textrm{disc}}}{L_{\textrm{Edd}}}\Bigg)$.  Using the accretion disc luminosity as opposed to the bolometric luminosity is acceptable since $L_{\textnormal{disc}}$ should dominate the bolometric luminosity.  The mean Eddington ratio corresponds to an Eddington rate of $\sim$40\%.  The uncertainty is rather large and dominated by the unknown obscurer geometry (cf. the broad model M contours in Figure \ref{fig:f6}), but these are robust uncertainties incorporating all systematics.  Furthermore, as we discuss in the next section, the implied luminosity is high even at the lower uncertainty limit, and is consistent with other multi-wavelength diagnostics.

To compare with a bolometric luminosity determined Eddington ratio, we use the bolometric correction factor of $\sim$10\,--\,30 from \citet{Vasudevan2010} for converting X-ray to bolometric luminosity.  This gives a slightly shifted range of Eddington ratios of $\textrm{log}_{10}(\lambda_{\textrm{Edd}})\,=\,-1.6$\,$\rightarrow$\,$0.6$, which corresponds to $\gtrsim$\,2.5\% of the Eddington rate (with the upper end being considerably super-Eddington).

\subsection{Comparison with other \textit{bona fide} CTK sources}
\label{sec:dis_BFcomparison}

\subsubsection{Ratio of intrinsic to observed luminosity}
\label{sec:BFratio}

The intrinsic parameters determined here with broad-band spectral fitting are consistent with multiple observations reported over almost two decades showing a lack of extreme variability in the source.  This allows us to stipulate IC 3639 as a \textit{bona fide} CTK source.  To date, there exists just $\sim$30 \textit{bona fide} CTK sources, names of which are collated in Table \ref{BF}.  Here, a \textit{bona fide} CTK source shows CTK column densities based on X-ray spectral analysis, and lacks extreme variability in the X-ray band.  The ID numbers presented in all bona fide CTK source plots herein correspond to the values shown in Table \ref{BF}.

\begin{figure}
\centering
\includegraphics[width=1\columnwidth]{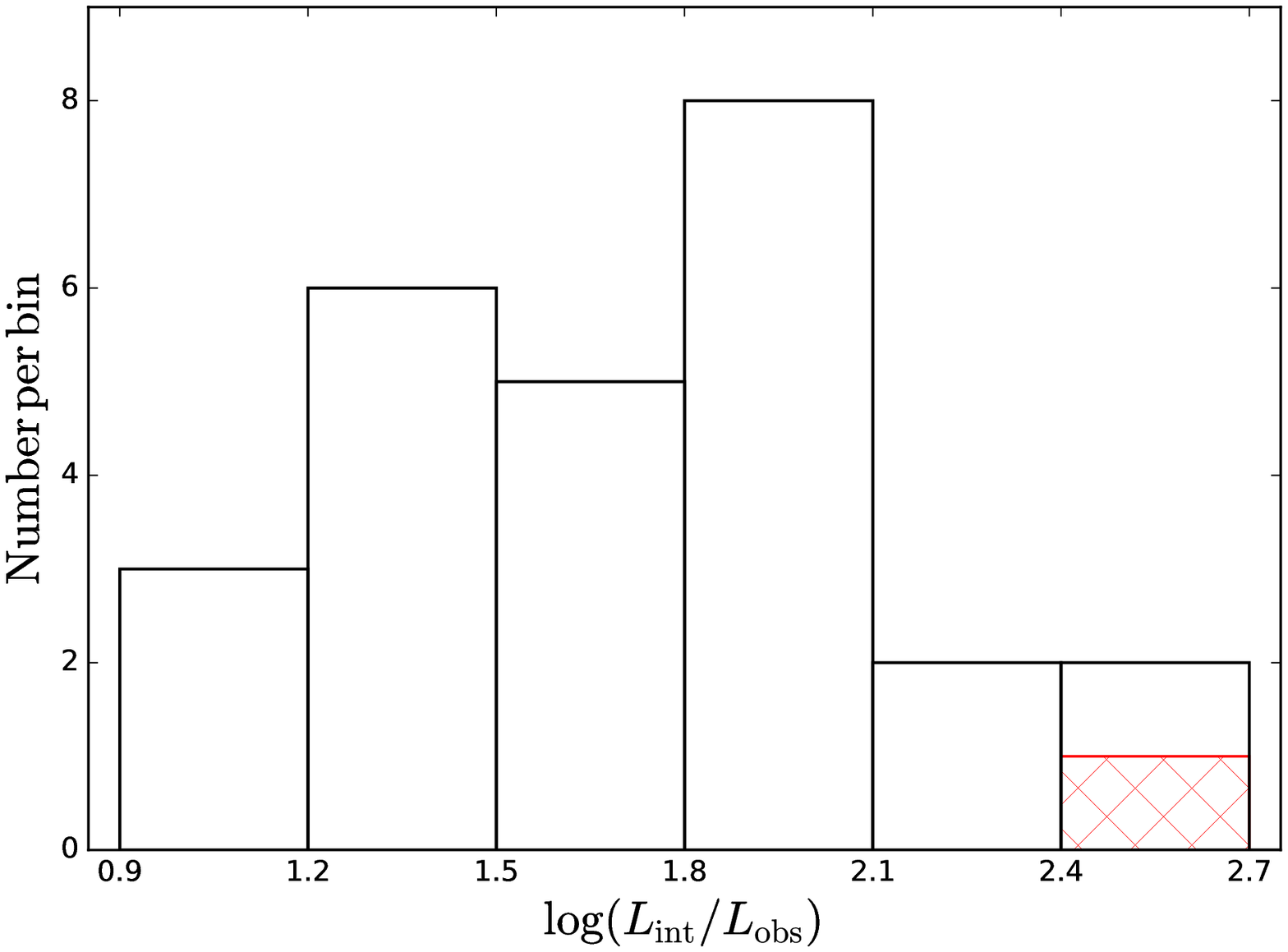}
\caption{Distribution of ratios of intrinsic to observed 2\,--\,10\,keV luminosity for the bona fide CTK AGN listed in Table \ref{BF}.  IC 3639 shows a comparatively large ratio, at $2.5^{+0.9}_{-1.3}$, and is represented as a red hatched patch in the distribution.  The other source in this bin is NGC 1068.}
\label{fig:f7}
\end{figure}

IC 3639 appears to show a comparatively high ratio of intrinsic to observed luminosity $\big(\sfrac{L_{\textrm{int}}}{L_{\textrm{obs}}}\big)$ relative to other bona fide CTK sources.  Here we again specify the X-ray luminosity in the 2\,--\,10\,keV band, and the intrinsic luminosity to be the absorption corrected luminosity.  Given the observed 2\,--\,10\,keV luminosity of $\textrm{log}_{10}(L_{\textrm{2-10\,keV}}^{\textrm{obs}}\,\textrm{[erg\,s}^{-1}])\,=\,40.79^{+0.04}_{-0.11}$, IC 3639 has:

$\textrm{log}_{10}\Bigg(\displaystyle \frac{L_{\textrm{int}}}{L_{\textrm{obs}}}\Bigg)=2.6^{+0.6}_{-1.1}$,

\noindent corresponding to a luminosity ratio of almost 300.  In comparison with the other bona fide sources listed in Table \ref{BF}, there exists just one other source with such a comparatively high ratio - NGC 1068.  The distribution of this ratio amongst the bona fide CTK AGN is shown in Figure \ref{fig:f7}.
Such a high value of the ratio complements the high column density predicted for the source based on multi-wavelength indicators, discussed next.

\subsubsection{Multi-wavelength indicators}
\label{sec:BFmulti_wavelength}

The large correction from observed to intrinsic X-ray luminosity for IC 3639 should be checked with independent methods, and for this we use multi-wavelength comparisons with the MIR and [OIII] luminosities.  Using the published value of the reddening corrected [OIII] flux for IC 3639 \citep{LaMassa2010}, we use a distance to the source of 53.6\,Mpc to calculate the [OIII] luminosity to be $\textrm{log}_{10}(L_{\textrm{[OIII]}}\,\textrm{[erg\,s}^{-1}])=42.0$.  Furthermore, the MIR (rest-frame 12$\mu m$) luminosity for IC 3639 is $\textrm{log}_{10}(L_{\textrm{MIR}}\,\textrm{[erg\,s}^{-1}])\,=\,43.52\pm0.04$ using high-angular resolution MIR imaging performed with ground-based 8-m class telescopes, providing subarcsecond resolution $\lesssim 0\farcs4$, corresponding to a physical resolution of $\lesssim 100\,$pc for IC 3639 \citep{Asmus2014,Asmus2015}.

\begin{figure*}
\center
\includegraphics[width=1\textwidth]{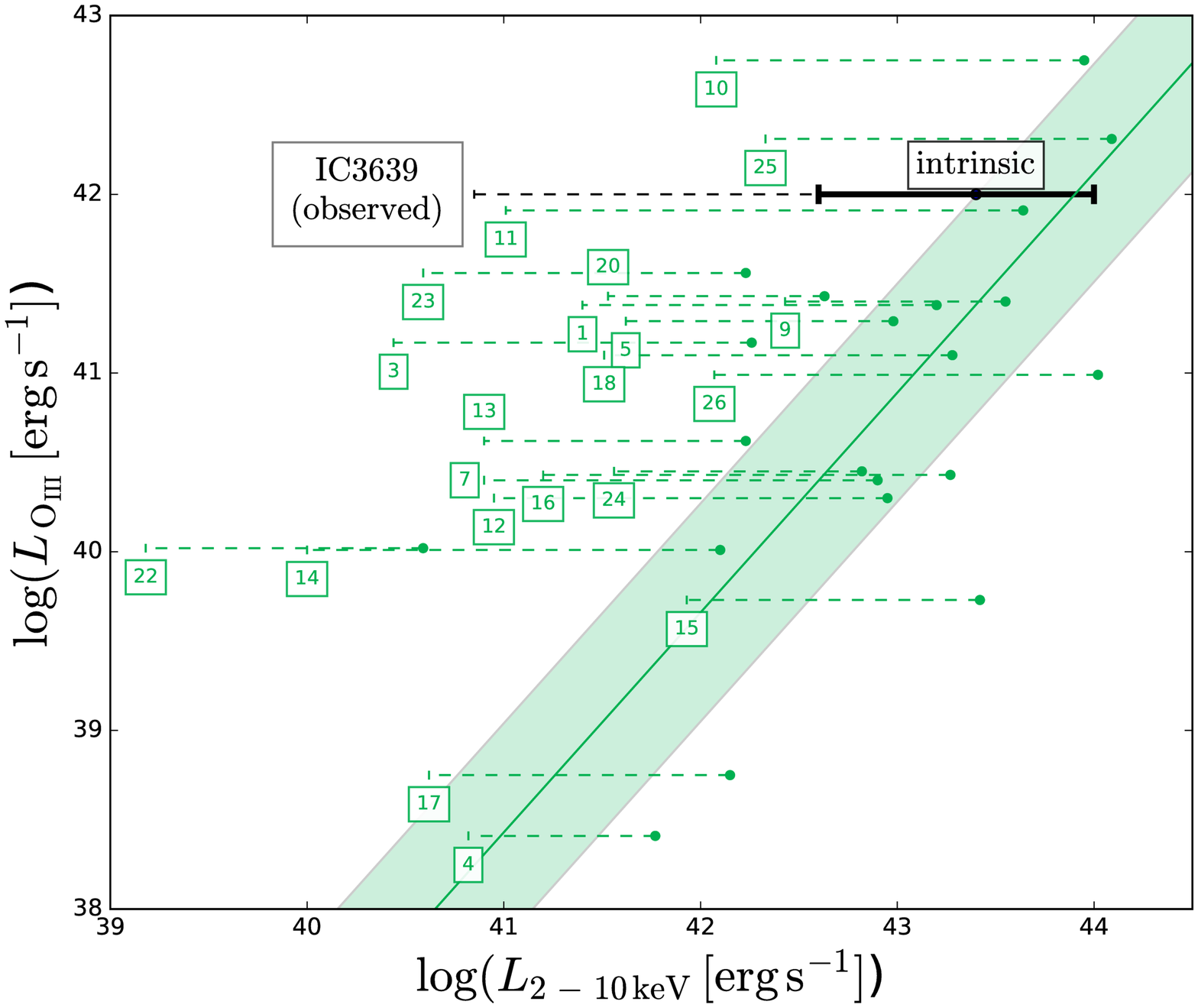}
\caption{Bona fide CTK AGN population overplotted on the $L_{\textrm{[OIII]}}$\,--\,$L_{\textrm{2\,--\,10\,keV}}$ relation from \citet{Berney2015}.  The shaded region represents the 1\,--\,$\sigma$ scatter.  Observed source X-ray luminosities before accounting for obscuration are shown as points to the left of horizontal bars and intrinsic values as points.  IC 3639 is shown in black, with uncertainty between models T and M for the X-ray emission represented as an error bar.  Note the error bar represents the luminosity derived to the 1\,--\,$\sigma$ level, $\textrm{log}_{10}(L_{\textrm{2-10\,keV}}\,\textrm{[erg\,s}^{-1}])\,=\,43.4^{+0.6}_{-0.8}$.  All source ID values are given in Table \ref{BF}.}
\label{fig:f8}
\end{figure*}

The [OIII] emission-line vs. X-ray luminosity relation from \citet{Berney2015} is presented in Figure \ref{fig:f8}, with the shaded region corresponding to the 1\,--\,$\sigma$ confidence level from the original study.  Over-plotted are all the \textit{bona fide} CTK sources from Table \ref{BF}.  This plot illustrates the effect of correctly modelling the obscuration surrounding the sources to give a better estimate of the intrinsic X-ray luminosity in the 2\,--\,10\,keV energy band.  Many of the sources have \textit{intrinsic} X-ray luminosities in better agreement with the relation, IC 3639 being an example.

\begin{figure*}
\center
\includegraphics[width=1\textwidth]{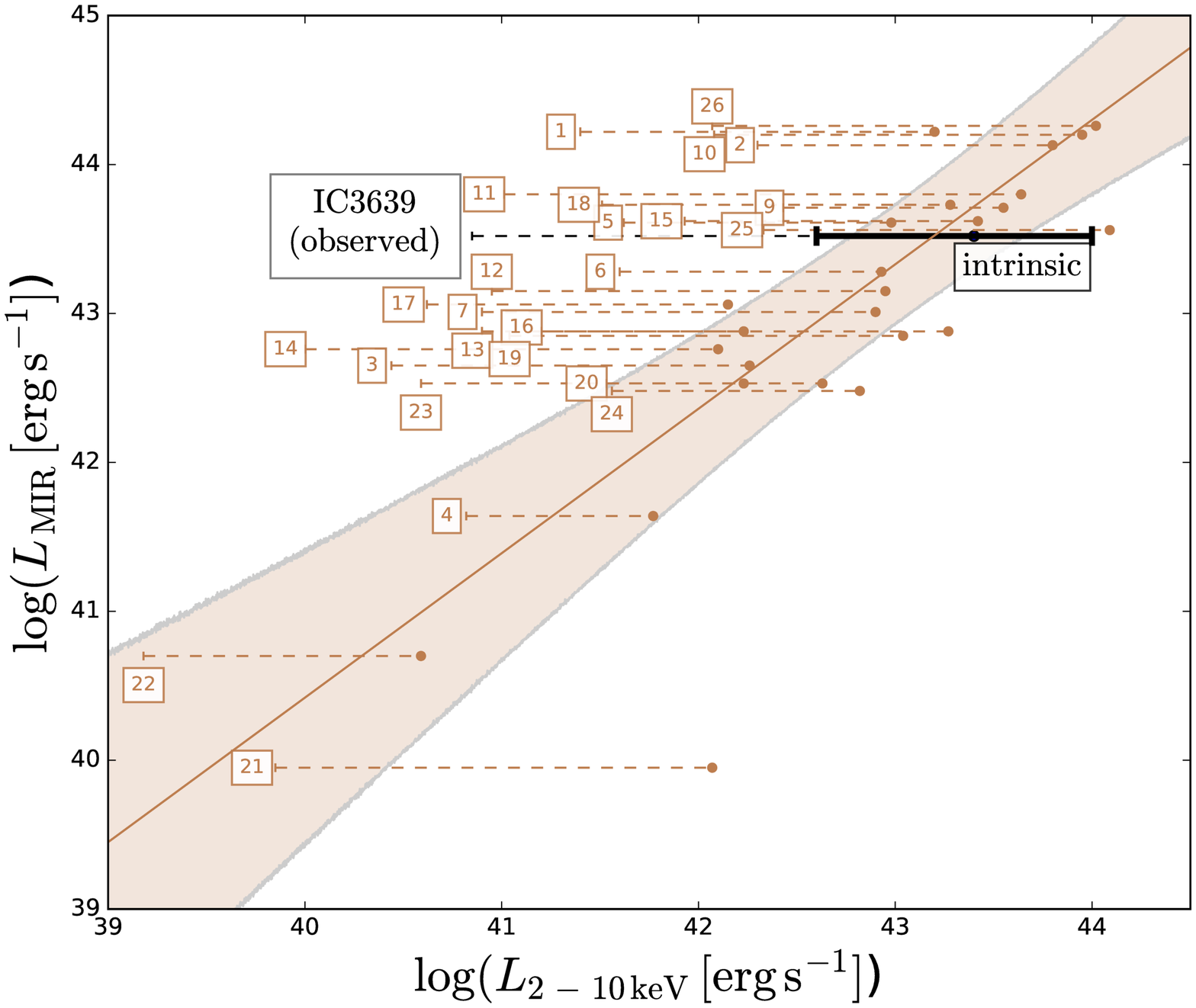}
\caption{Bona fide CTK AGN population overplotted on the $L_{\textrm{MIR}}$\,--\,$L_{\textrm{2\,--\,10\,keV}}$ relation from \citet{Asmus2015}.  The shaded region represents the 1\,--\,$\sigma$ scatter generated from Monte Carlo modelling of the relation uncertainties presented in the original paper.  Observed source X-ray luminosities before correcting for obscuration are represented as points to the left of horizontal bars and intrinsic values as points.  IC 3639 is shown in black, with uncertainty between models T and M for the X-ray emission represented as an error bar.  Note the error bar represents the luminosity derived to the 1\,--\,$\sigma$ level, $\textrm{log}_{10}(L_{\textrm{2-10\,keV}}\,\textrm{[erg\,s}^{-1}])\,=\,43.4^{+0.6}_{-0.8}$.  All source ID values are given in Table \ref{BF}.}
\label{fig:f9}
\end{figure*}

We also reproduce the relation between intrinsic X-ray luminosity and MIR luminosity from \citet{Asmus2015} in Figure \ref{fig:f9}.  The shaded region shows the 1\,--\,$\sigma$ confidence region generated through Monte-Carlo generated uncertainties from uncertainties determined in the study.  The MIR luminosities of bona fide CTK sources were either accumulated from \citet{Asmus2015} or from the \textit{Wide-field Infrared Survey Explorer (WISE)} all-sky survey\footnote{http://irsa.ipac.caltech.edu/cgi-bin/Gator/nph-dd}, both calculated at 12\,$\mu$m.  Again, many bona fide CTK sources, including IC 3639, show improved agreement with the relation.  The exception is NGC 4945, which has been scrutinised to explain its nature: for example, \citet{Puccetti2014} suggest most of the high-energy emission is transmitted rather than scattered, whereas \citet{Brightman2015} suggests the source to have a high covering factor.  See \citet{Gandhi2015} and \citet{Asmus2015} for further discussion of recent studies of NGC 4945.  

The \citeauthor{Berney2015} relation gives a predicted X-ray luminosity of $\textrm{log}_{10}(L_{\textrm{[2\,--\,10\,keV]}}\,\textrm{[erg\,s}^{-1}])\sim43.9\pm2.4$, whereas the \citeauthor{Asmus2015} relation gives a predicted X-ray luminosity for IC 3639 of $\textrm{log}_{10}(L_{\textrm{[2\,--\,10\,keV]}}\,\textrm{[erg\,s}^{-1}])\,=\,43.17\pm0.37$ \citep{Asmus2015}.  Thus both \textit{predicted} intrinsic X-ray luminosities are fully consistent with the directly modelled 2\,--\,10\,keV luminosity that we derive for IC 3639.

\subsubsection{The Fe-K$\alpha$ fluorescence line and the Future}
\label{sec:BFironKa}

A high EW is indicative of strong reflection within a source, as detailed in Section \ref{sec:introduction}.  However, across the full spectrum of known bona fide CTK sources, there are a broad range of EWs, including values less than 1\,keV.  The lowest EW value determined to date for a bona fide source is reported by \citet{Gandhi2016} for NGC 7674, with an Fe-K$\alpha$ line EW of 0.38$_{-0.09}^{+0.10}$\,keV.
Figure \ref{fig:f10} compares the Fe-K$\alpha$ strength relative to a power law continuum for IC 3639 and NGC 7674. The data are from the combined \textit{Suzaku} XIS03 detectors for both.  IC 3639 shows a peak of the Fe-K$\alpha$ line consistent with ten times that of the continuum model, whereas NGC 7674 shows a peak of the Fe-K$\alpha$ line around two times the corresponding continuum power law model.  This illustrates the broad range in EWs, as well as the need for improved diagnostics used to confirm candidate CTK AGN.

A large EW could correlate with the large SFRs found here.  \citet{Levenson2002} suggest that the mechanical energy provided through periods of strong star formation could effectively \textit{inflate} the torus, altering the covering factor and thus EW associated with the Fe-K$\alpha$ fluorescence line.

\begin{subfigures}
\begin{figure}
\centering
\includegraphics[width=0.8\columnwidth,angle=-90,trim={35cm 30cm 0cm 0cm}]{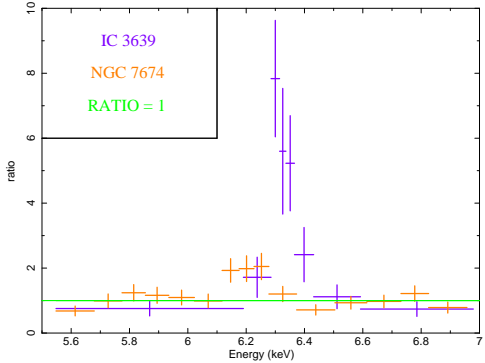}
\caption{Graph of ratio between model and observed counts against energy.  The model used was a simple power law to model the continuum of the source.  Purple and orange data sets show the comparative strength of the iron line relative to the continuum for IC 3639 and NGC 7674 respectively.  The horizontal green line represents a ratio of unity.}
\label{fig:f10}
\end{figure}
\begin{figure}
\center
\includegraphics[width=0.6\columnwidth,angle=-90]{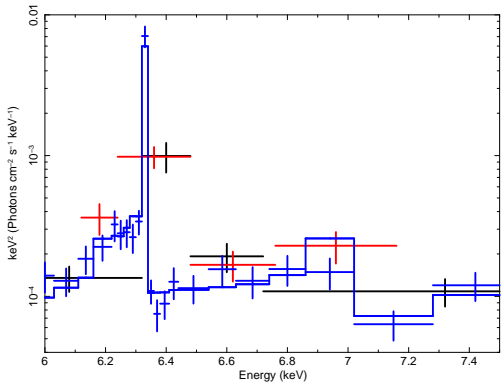}
\caption{Simulated 100\,ks \textit{Athena} spectrum (shown in blue), together with the model used to generate the data.  \textit{NuSTAR} FPMA/FPMB data points and model are superimposed in black/red respectively.  The \textit{Athena} data has been re-binned for clarity.  The Fe-K$\alpha$ line complex is clearly resolved to display additional components, such as the Compton shoulder.}
\label{fig:f11}
\end{figure}
\end{subfigures}

\begin{table}
\centering
\caption{IDs corresponding to all currently known \textit{bona fide} CTK AGN, including IC 3639, in reference to Figures \ref{fig:f7}, \ref{fig:f8} and \ref{fig:f9}.}
\label{BF}
\begin{tabular}{rlcrlcrl}
\hline \hline
ID & Name 		 && ID & Name 		 && ID & Name 		 \\
\hline \hline
1  & Arp 299B    && 11 & NGC 1068    && 21 & NGC 4945    \\
2  & CGC G420-15 && 12 & NGC 1320    && 22 & NGC 5194    \\
3  & Circinus    && 13 & NGC 2273    && 23 & NGC 5643    \\
4  & ESO 005-G004&& 14 & NGC 3079    && 24 & NGC 5728    \\
5  & ESO 138-G001&& 15 & NGC 3281    && 25 & NGC 6240S   \\
6  & ESO 565-G019&& 16 & NGC 3393    && 26 & NGC 7674    \\
7  & IC 2560     && 17 & NGC 4102    &&	   &			 \\
8  & IC 3639     && 18 & NGC 424     &&	   &			 \\
9  & Mrk 3       && 19 & NGC 4785    &&	   &			 \\
10 & Mrk 34      && 20 & NGC 4939    &&	   &		 	 \\
\hline \hline
\end{tabular}
\end{table}

Future missions such as \textit{Athena} \citep{Nandra2013} hold the potential to resolve fluorescence complexes in much greater detail.  In particular, resolved spectral imaging of the Compton shoulder could tell us more about how buried IC 3639 is in the surrounding obscuring shroud of dust.  Figure \ref{fig:f11} illustrates simulated data for the proposed \textit{Athena} X-ray Integral Field Unit (XIFU), which will have a spectral resolution of $\sim$2.5\,eV at 6\,keV.  We used the response and background files provided by the \textit{Athena} website\footnote{http://x-ifu-resources.irap.omp.eu/PUBLIC/BACKGROUND/5arcsec/} together with an exposure of 100\,ks.  Over-plotted are the equivalent \textit{NuSTAR} FPMA and FPMB data points from this work for the same region fitted with model T.  A clear detection of the Compton shoulder and other fluorescence lines are visible with the \textit{Athena} spectra, and could be used to investigate super-Solar abundances for IC 3639 in greater detail due to the higher SNR predicted (the current simulation assumes Solar abundances).

\section{SUMMARY}
\label{sec:conclusions}
Recent \textit{NuSTAR} observations were combined with archival \textit{Suzaku} observations of the nearby type 2 Seyfert AGN IC 3639.  Our key findings are enumerated below.

\begin{description}
\item[1 ] We used the \textsc{mytorus} and \textsc{bntorus} models to self-consistently fit the broadband spectral data available for IC 3639.  These predominantly show a very high level of obscuration, favouring column densities of order $N_\textrm{H}\,\sim\,1.0 \times 10^{\textrm{25}}\,\textrm{cm}^{\textrm{-2}}$.  This is consistent with previous results from the literature, suggesting a lack of variability over the past two decades between the \textit{BeppoSAX} and \textit{NuSTAR} observations.  As a result, we classify IC 3639 as a \textit{bona fide} CTK AGN.\\

\item[2 ] We consider the \textit{Suzaku} HXD observation of the source to be a non-detection after accounting for the high background level and its reproducibility.  This contradicts a previous study of the same HXD data set.\\

\item[3 ] The combined results of the two torus models give an intrinsic X-ray luminosity (2\,--\,10\,keV band) of $\textrm{log}_{10}(L_{\textrm{2-10\,keV}}\,\textrm{[erg\,s}^{-1}])\,=\,43.4^{+0.6}_{-1.1}$.  We then predict a source Eddington ratio of $\textrm{log}_{10}(\lambda_{\textrm{Edd}})=-0.4^{+0.8}_{-1.1}$, to the 1\,--\,$\sigma$ confidence level.\\

\item[4 ] We find an extreme EW of the Fe-K$\alpha$ fluorescence line for the source of $2.94^{+2.79}_{-1.30}$\,keV, consistent with \citet{Risaliti1999}, and one of the highest amongst bona fide CTK AGN.  The source also shows a high intrinsic to observed 2\,--\,10\,keV luminosity ratio.\\

\item[5 ] A multi-wavelength comparison between X-ray and MIR continuum and [OIII] emission line fluxes of IC 3639 with all known \textit{bona fide} CTK AGN give good agreement with known intrinsic correlations.  This provides independent evidence that we are robustly measuring the absorption-corrected X-ray luminosity.\\
\end{description}

Further studies of other local CTK candidates are clearly vital to properly ascertain the cosmological processes behind the formation of different AGN classes, as well as to help resolve the peak of the CXB flux.

\section*{ACKNOWLEDGEMENTS}
\label{sec:acknowledgements}
This work made use of data from the {\it NuSTAR} mission, a project led by the California Institute of Technology, managed by the Jet Propulsion Laboratory, and funded by the National Aeronautics and Space Administration. We thank the {\it NuSTAR} Operations, Software and Calibration teams for support with the execution and analysis of these observations.  This research has made use of the {\it NuSTAR} Data Analysis Software (NuSTARDAS) jointly developed by the ASI Science Data Center (ASDC, Italy) and the California Institute of Technology (USA).

\textit{Facilities: NuSTAR, Suzaku, Chandra}.

We thank the anonymous referee for their invaluable comments which have helped to improve the paper.

The scientific results reported in this article are based on observations made by the \textit{Chandra} X-ray Observatory

This research has made use of data, software and/or web tools obtained from the High Energy Astrophysics Science Archive Research Center (HEASARC), a service of the Astrophysics Science Division at NASA/GSFC and of the Smithsonian Astrophysical Observatory's High Energy Astrophysics Division.

This research has made use of the NASA/IPAC Extragalactic Database (NED), which is operated by the Jet Propulsion Laboratory, California Institute of Technology, under contract with the National Aeronautics and Space Administration.

This research has made use of data obtained from the Suzaku satellite, a collaborative mission between the space agencies of Japan (JAXA) and the USA (NASA).

This publication makes use of data products from the Wide-field Infrared Survey Explorer, which is a joint project of the University of California, Los Angeles, and the Jet Propulsion Laboratory/California Institute of Technology, funded by the National Aeronautics and Space Administration.

P.B. thanks STFC and the RAS for funding.

P.G. thanks STFC for support (grant reference ST/J003697/2).

A.A. acknowledges financial support from Majlis Amanah Rakyat (MARA), Malaysia.

D.A. acknowledges the Science and Technology Facilities Council (DMA; ST/L00075X/1).

W.N.B. acknowledges Caltech NuSTAR subcontract 44A-1092750 and the V.M. Willaman Endowment.

S.F.H. acknowledges support from the European Research Council under Horizon 2020 grant ERC-2015-StG-677117.

M.\,K. acknowledges support from the Swiss National Science Foundation (SNSF) through the Ambizione fellowship grant PZ00P2\textunderscore154799/1

S.M.L. acknowledges support by an appointment to the NASA Postdoctoral Program at the NASA Goddard Space Flight Center, administered by Universities Space Research Association under contract with NASA.

A.M. acknowledges support from the ASI/INAF grant I/037/12/0-011/13.

F.E.B. and C.R. acknowledge support from NASA NuSTAR A01 Award NNX15AV27G, CONICYT-Chile grants Basal-CATA PFB-06/2007, FONDECYT Regular 1141218 and 1151408, "EMBIGGEN" Anillo ACT1101, the China-CONICYT, and the Ministry of Economy, Development, and Tourism's Millennium Science Initiative through grant IC120009, awarded to The Millennium Institute
of Astrophysics, MAS.

\appendix
\section{ADDITIONAL CONTOUR PLOTS}

\begin{figure*}[!tbp]
  \centering
  \begin{minipage}[b]{0.4\textwidth}
    \includegraphics[width=0.8\textwidth,angle=-90,trim={0cm 3cm 0cm 0cm}]{fA1.eps}
  \end{minipage}
  \hfill
  \begin{minipage}[b]{0.4\textwidth}
    \includegraphics[width=0.8\textwidth,angle=-90,trim={2.5cm 5cm 1.5cm 0cm}]{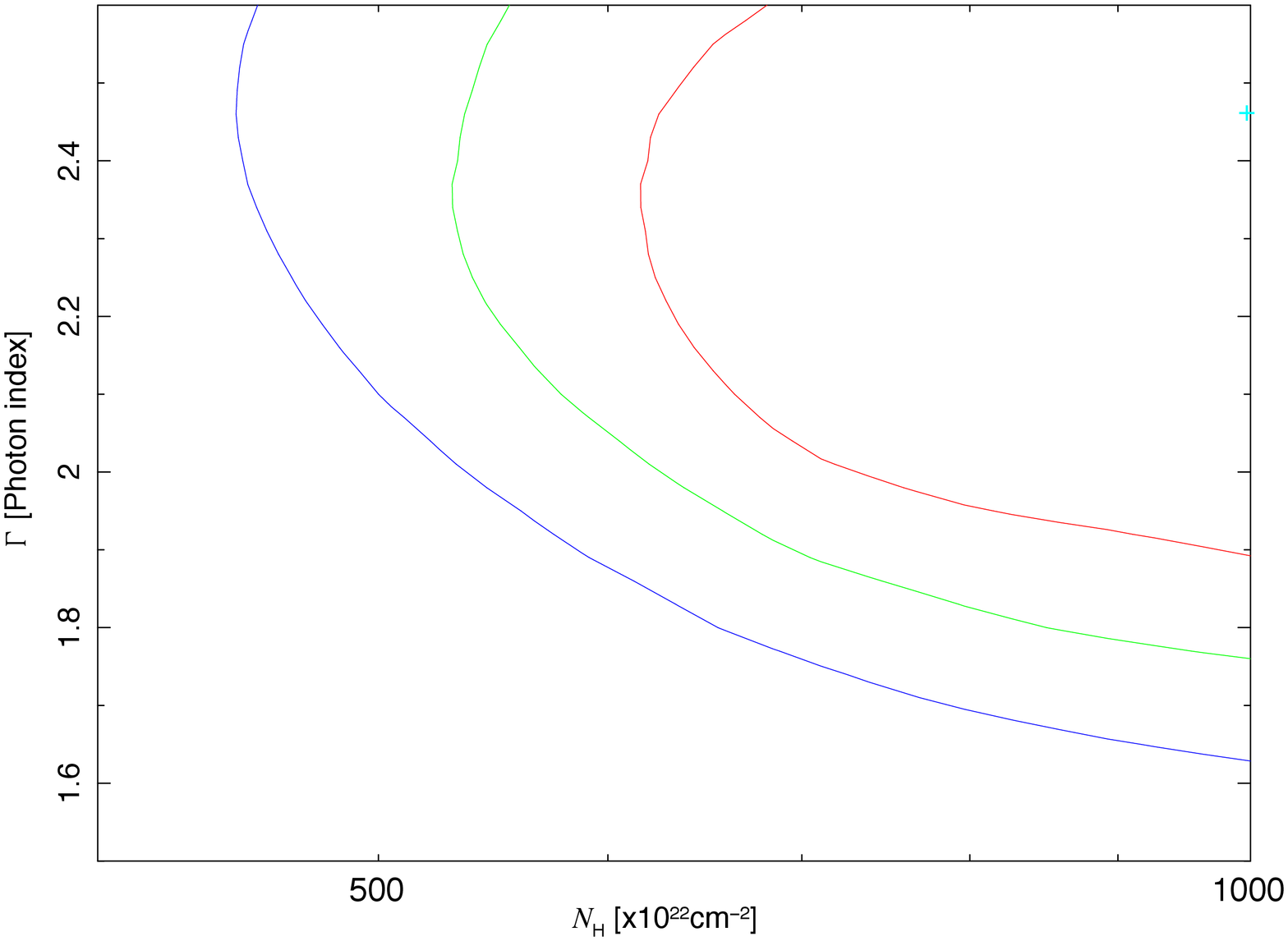}
  \end{minipage}
  \caption{\label{fig:fA1} Contour plots for the \textsc{bntorus} model (model T -- left panel), and \textsc{mytorus} model (model M -- right panel), between power law photon index and obscuring column density.  The model T contour shows the LOS obscuring column density, whereas model M shows the equatorial column density.  The LOS column is CTK to 99\% confidence for both models.  In addition, the plots illustrate the unconstrained nature of the obscuring column, even beyond the model maximum of $N_\textrm{H}\,=\,1.0\,\times\,10^{26}\,\textrm{cm}^{-2}$ for model T.  The red, green and blue contours represent the 1\,--\,$\sigma$, 90\% and 99\% confidence levels, respectively.  The cyan cross represents the best-fit values found with \textsc{xspec}.}
\end{figure*}

Here we include the contour plots between photon index and column density for both toroidal models T and M.  The column density plotted for model T (Figure \ref{fig:fA1}, left panel) corresponds to the LOS column density, whereas the column used in the model M contour plot (Figure \ref{fig:fA1}, right panel) corresponds to the equatorial column density.   To 99\% confidence (blue contour line), the corresponding LOS obscuring column density for both models is $\gtrsim\,4\,\times\,10^{24}\,\mathrm{cm}^{-2}$ - well into the CTK regime, and is unconstrained at the upper end allowed by either model.  Although the model T contour plot illustrates a wider range in parameter space than the model M contour, the values found are clearly consistent between the two graphs.

\bibliography{./bibliography}

\listofchanges

\end{document}